\documentclass[aps,prb,preprint,showpacs,showkeys,superscriptaddress,groupedaddress]{revtex4-1}
\usepackage{graphicx}  
\usepackage{dcolumn}   
\usepackage{bm}        
\usepackage{amssymb}   
\usepackage{color,soul}
\usepackage{amsmath}
\begin{document} 
\title{Evolutionary optimization of PAW data-sets for accurate high pressure simulations}
\author{Kanchan~Sarkar} \affiliation{Department of Chemical Engineering and Materials Science, University of Minnesota, Minneapolis, MN 55455, USA}
\author{Mehmet~Topsakal} \affiliation{Department of Chemical Engineering and Materials Science, University of Minnesota, Minneapolis, MN 55455, USA}
\author{N. A. W. Holzwarth} \affiliation{Department of Physics, Wake Forest University, Winston-Salem, NC 27109 USA}
\author{and Renata~M.~Wentzcovitch} \affiliation{Department of Chemical Engineering and Materials Science, University of Minnesota, Minneapolis, MN 55455, USA}
\affiliation{Minnesota Supercomputing Institute for Digital Technology and Advanced Computations, University of Minnesota, Minneapolis, MN 55455, USA}
\date{\today}
\begin{abstract}
We examine the challenge of performing accurate electronic structure calculations at high pressures by comparing the results of all-electron full potential linearized augmented-plane-wave calculations with those of the projector augmented wave (PAW) method. In particular, we focus on developing an automated and consistent way of generating transferable PAW data-sets that can closely produce the all electron equation of state defined from zero to arbitrary high pressures. The technique we propose is an evolutionary search procedure that exploits the ATOMPAW code to generate atomic data-sets and the Quantum ESPRESSO software suite for total energy calculations. We demonstrate different aspects of its workability by optimizing PAW basis functions of some elements relatively abundant in planetary interiors. In addition, we introduce a new measure of atomic data-set goodness by considering their performance uniformity over an enlarged pressure range.
\end{abstract}

\pacs{02.60.Pn, 63.20.dk, 71.15.Ap, 91.60.Fe, 91.60.Gf}
\keywords{PAW data-sets, high pressure simulation, evolutionary computing, goodness measure of data-set performance, genetic algorithms}
\maketitle

\section{Introduction}\label{int}  

Exploring material properties at extreme conditions is key for modeling planetary interiors\cite{Umemoto2006, Militzer2008, Martinez2012,  Smith2014}. These properties at ultra-high pressure and temperature conditions control the dynamical evolution of planets\cite{Yuen1,Tackley}, and provide important inputs for geodynamics simulations. They are also indispensable for interpretation of Earth’s seismic tomography models\cite{wentz06,WuWen14}. In this context, {\em{ab initio}} calculations based on Density Functional Theory (DFT)\cite{Hohenberg1964, Kohn1965} have proved fundamental in predicting material properties at extreme conditions.\cite{RIMG2010}

There is a wide variety of available numerical implementations of DFT. The quantitative reproducibility of major solid-state DFT codes around zero pressure has recently been analyzed by Lejaeghere {\em{et al.}}\cite{Lejaeghereaad3000,Lejaeghere2014} in terms of a ``$\Delta$-factor'' criterion for the elemental materials throughout the Periodic Table. All of the numerical schemes rely on parameters tuned to each atom and method. For example, the pseudopotential method uses a cut-off radii as adjustable parameters to generate psedopotential "data-sets" used in solid state calculations. Libraries of atomic data-sets such as GBRV,\cite{Garrity2014} JTH,\cite{Jollet2014} pslibrary\cite{dal_corso_pseudopotentials_2014} for describing materials near equilibrium are available for many of the numerical schemes and computer codes.

The study of materials under high pressure requires additional numerical considerations. The all-electron full-potential linearized augmented-plane-wave (AE-FLAPW) method\cite{Yu91} such as implemented in the WIEN2k code\cite{wien2k:2001} is able to treat high pressure materials provided that muffin-tin radii and convergence parameters are adjusted appropriately. In this work, we take the WIEN2k results as our target reference for optimizing atomic data-sets for high-pressure studies. Here, we use the projector augmented wave method\cite{Blochl1994} (PAW), which takes advantage of the numerical efficiency of pseudopotential-like formalisms while retaining the accuracy of all-electron treatments. By constructing high-quality atomic PAW data-sets, it has been shown that excellent agreement between PAW and WIEN2k results can be achieved. However, extending the capabilities of PAW calculations to high pressure simulations, generally requires adjustments to the atomic PAW data-sets due to the presence of several computational and numerical challenges such as presence of unphysical solutions (ghost states\cite{gonze:1990}), alteration of the optimal augmentation radii for computational efficiency, promotion of semi-core electrons to the valence\cite{holzwarth:2001, Tackett2001} {\em{etc}}.

The construction and testing of robust and soft atomic data-sets for accurate high pressure simulations is generally a time-consuming and patience-challenging task, demanding careful supervision during the simultaneous optimization of accuracy and computational expense. In this regard, evolutionary computing (EC) techniques\cite{Goldberg1989,Michalewicz97} can offer powerful alternative tools to find such optimal PAW data-set and to minimize, if not eliminate altogether such user supervisions.

Genetic Algorithms (GA),\cite{Goldberg1989, Michalewicz97, Bennett1994, Wu2014, Sarkar2015} the most well-known members of EC family, have found relevance in virtually all fields of scientific and technological applications. GAs start by building a population of plausible solutions that are given a chromosomal representation and defining a fitness function. The fitness function produces a numeric score to measure the degree of acceptability of a solution (or individual) being proposed. The individuals in the population tend to evolve through generations ({\em{i.e.}} iterations) towards higher fitness in the fitness landscape (or energy landscape) under suitable genetic operations like selection, crossover, mutation, {\em{etc.}} Selection process enforces the Darwinian Principle of the survival of the fittest, probabilistically favoring individuals with higher quality to become parents for the next generation. Crossover and mutation cause small random unbiased changes to the individuals in a population. More specifically, the crossover operator brings in more viable parts of two chromosomal solutions onto the same member while mutation introduces features that were lost or missing in the current population.

To the best of our knowledge, the first successful examples of evolutionary computing to generate atomic data-sets were reported by Brock {\em{et al.}}\cite{brock2016} and Hansel {\em{et al.}}\cite{hansel2015} who used a multi-objective GA to automate a search for Pareto optimal set of pseudopotentials with varying user selectable compromises between accuracy and efficiency. In this contribution we develop an evolutionary computing guided recipe for optimizing PAW atomic data-sets over a wide pressure range. The problem has been formulated as a single objective constrained optimization procedure. 

In section \ref{formulation}, we represent the formulation of the optimization approach to shape PAW atomic data-sets along with brief descriptions of useful tunable options in the ATOMPAW code. Section \ref{ethodology} presents the methodologies we have proposed to generate PAW data-sets {\em{uniformly}} optimized up to very high pressures. Section \ref{Comp} shows details of DFT calculations that have been carried out in this work. Next (section \ref{RD}) we investigate the workability of the proposed soft-computing strategy by optimizing PAW basis functions for some important planet forming elements including carbon, magnesium, aluminum, silicon, and iron. In section \ref{GDP}, we present an improved measure of ``goodness" of atomic data-sets and compare the new measure with some existing measures. Our conclusions are summarized in section \ref{concl}.

\section{Formulation \label{formulation} }

The Projector Augmented Wave (PAW) formalism\cite{Blochl1994} uses a number of parameters, radial functions representing orbital bases and projectors, localized charge moments, and the local pseudo potential. In the present work, we use the ATOMPAW code\cite{holzwarth:2001} to generate atomic data-sets and the Quantum ESPRESSO\cite{PWscf:2009} and WIEN2k software\cite{wien2k:2001} to produce equations of state for the optimization process. Over the years, a number of options for constructing atomic PAW datasets have accumulated in the ATOMPAW code. Some of the most useful options {\em{i.e.}}, radial matching parameters, energies, and functional shapes are documented in Appendix (\ref{aapndx}). There is considerable flexibility in tuning those options to strike the right balance between desired accuracy and efficiency in the course of constructing atomic data-sets. The purpose is to reproduce the all-electron behavior of each atom as accurately and efficiently as possible to arbitrary pressures. Operationally, there are two types of procedures involved with the optimization of the PAW data-set: one at the atomic level that sets up a trial data-set and another one that evaluates the data-set ability to represent the behavior that the atom in the solid.

The atomic procedure involves the calculation of the electronic structure of an atom in which the all-electron (AE) and pseudo (PS) functions are setup. As discussed in more detail in the Appendix (\ref{aapndx}), there are several schemes for constructing the PS functions. Within a given scheme, there are a number ($M$) of adjustable variables which we will denote as an array $s$ (analogous to a solution string or individual in genetic algorithms) where
\begin{equation} \label{stringdef} 
s \equiv \{v_1, v_2, \dots, v_M\}. 
\end{equation} 
The variables $v_j$ generally represent various matching radii and basis function energies. For example, a typical input to the ATOMPAW program is shown in Fig. \ref{fig:table} with descriptive commentary. Thus the string ($s^C$) for carbon is
\begin{eqnarray} 
s^{C} = \{r_c, r_{\rm shape}, r_{\rm 
vloc}, r_{\rm core}, r_{c1}, r_{c2}, r_{c3}, r_{c4}, E_{\rm ref1}, 
E_{\rm ref2}\} 
\end{eqnarray} 
The meaning of these parameters are explained in Fig. \ref{fig:table} and Appendix \ref{aapndx}. The value of each parameter in the string $s$ can vary within a certain range specified by:
\begin{equation} 
\label{range}
 v_{j}^{ \rm min} \le v_j \le v_{j}^{ \rm max}. 
\end{equation} 
In addition to the constraints on the range of parameter values, there is an optimization condition on each variable in set $s$ imposed by the behavior of the logarithmic derivative curves of the all-electron and pseudo radial wavefunctions, which should be as similar as possible within a certain energy range. A measure of the accuracy of the logarithmic derivatives for any given variable set $s$ can be defined as follows 
\begin{equation} \label{logd} 
O^{s}_{atom}=\sqrt{\sum_{l=0}^{l_{max}}\sum_{E} \left[ 
d_{E}^{l}(\text{PS}^{s})-d_{E}^{l}(\text{AE}) \right]^2}, 
\end{equation}
where $l$ is the angular momentum quantum number for an atomic orbital and $l_{max}$ is the maximum orbital angular momentum needed. In Eq. (\ref{logd}) the logarithmic derivatives, $d^l_E$, for the radial solution of the Kohn-Sham equation of the all-electron (AE) system and of the pseudized (PS$^s$) system using the variable set $s$ are evaluated at the augmentation radius $r_c$ and energy $E$, for $E$ values defined on a regularly spaced grid within a predefined range.

The second part of the procedure involves solid state electronic structure calculations to assess the performance of the set $s$ in representing the atom in the solid state environment. The goal is to find the optimized PAW parameters set ($s^{\star}$) that best describes the high-pressure cohesive and structural properties of solid containing the atom. In general, the target all-electron total energy versus volume curve needs to be reproduced, as closely as possible using the optimum data-set specified by $s^{\star}$. For quantitative comparisons it is convenient to fit the total energy versus volume curve produced by a data-set $s$ to a finite strain expansion. For the pressure range investigated here, the third order expansion, i.e., the Birch-Murnaghan equation of state\cite{Murnaghan1944, Birch1947} is adequate:
\begin{equation} \label{BM} 
E^s(V)=\frac{9}{16}B^s_{0}V^s_{0}\left \{ \left [ \left 
(\frac{V^s_{0}}{V} \right )^{2/3}-1 \right ]^{3} B'^s_0+ \left[ 
\left(\frac{V^s_{0}}{V} \right )^{2/3}-1\right ]^{2}\left [ 6-4 
\left(\frac{V^s_{0}}{V} \right)^{2/3} \right ] \right \} 
\end{equation} 
with the corresponding pressure versus volume relationship: 
\begin{equation}\label{PBM} P^s(V)= 
\frac{3 B^s_0}{2} \left[\left(\frac{V^s_{0}}{V} \right)^{7/3} - 
\left(\frac{V^s_{0}}{V} \right)^{5/3} \right ] \left\{1+ 
\frac{3}{4}(B'^s_0-4) \left[ \left(\frac{V^s_{0}}{V}\right)^{2/3} -1 
\right] \right\}, 
\end{equation} 
Here we have defined the zero of the energy as the energy at zero pressure. The equation of state parameters are: $V^s_0$, equilibrium volume, $B^s_0$, the zero pressure bulk modulus, and $B'^s_0$, its pressure derivative. They are determined from a least squares fit of the calculated total energies evaluated at several volumes to Eq. (\ref{BM}). The corresponding parameters for the all-electron equation of state, $V^{AE}_0$, $B^{AE}_0$, and $B'^{AE}_0$, are determined in a consistent manner.

The problem of generating PAW atomic files for an arbitrary pressure range can be cast into an equivalent optimization problem: the difference between the equation of state curves ($E(V)$ or $P(V)$) generated by AE and the PAW data-set generated by $s$ (subject to optimization) must be minimized within the entire pressure range under consideration. In addition, it is desirable to have uniform performance within the same pressure range. In practice, we may use the following objective function to accomplish this:
\begin{eqnarray} \label{obj} O^s_{\rm solid} &=& 
\frac{1}{n}\sum_{i=1}^{n} \omega _i \vert \Delta P^s(V_i) \vert 
\nonumber \\ \text{where } \vert \Delta P^s(V_i) \vert &=& \vert 
P^{s}(V_i)- P^{\text{AE}}(V_i) \vert. 
\end{eqnarray} 
Here $n$ is a number of closely spaced equidistant volume points ($V_i$) at which the pressure has been evaluated according to Eq. (\ref{PBM}). The reason for choosing pressure over energy is to accentuate the differences between the AE result and data-set result produced by string $s$. The weight factors $\omega _i$ in Eq. (\ref{obj}) can be used to optionally give special attention to a preferred pressure region. Therefore, the optimization problem involves setting and tuning a vector of free parameters ($s=r_c$, $r_{shape}$, $r_{vloc}$, $r_{core}$, $r_{c1}$, $r_{c2}$, $r_{c3}$, $r_{c4}$, $...$, $E_{ref1}$, $E_{ref2}$, $...$), the PAW data-set generator, to minimize the difference between AE and PAW log derivatives (Eq. \ref{logd}) and difference between equations of state (Eq. \ref{obj}), subject to the satisfaction of the constraints of Eq. \ref{range} plus a few additional constraints detailed in Sec. \ref{Comp}. Essentially this is a case of constrained non-linear optimization problem requiring a proper global optimization technique to strike the right balance between computational expense and the accuracy demanded from the optimized data-set.

Over the years, there has been a growing interest in the use of evolutionary computing methodologies in optimization problems owing to their ability to exploit the accumulated information about an initially unknown search space and bias subsequent searches into useful subspaces.\cite{Herrera1998}  In any evolutionary computing algorithm, choosing a proper fitness function is instrumental to explore a large search space more effectively and efficiently. The fitness function evaluates the goodness of a genetic string (or solution string or individual) in a population. In our case we have two fitness functions: one corresponding to the dataset performance in the atom and another corresponding to its performance in the solid. The fitness functions can be written in terms of the objective measures $O^s_{\rm atom}$, defined in Eq. \ref{logd}, and $O^s_{\rm solid}$, defined in Eq. \ref{obj}.
\begin{eqnarray} 
f^s=e^{-\lambda O^s }.\label{eqfit} 
\end{eqnarray}
where $\lambda$ is a small constant ($\sim 10^{-1}$ for a good initial guess $s$) that takes care of exponential underflow in case the objective function $O(s)$ of the solution string $s$ has a large value. The solution strings tend to get refined over generations by maximizing their fitness values.

We have employed two different evolutionary computing methodologies: a GA algorithm (left side in Fig.\ref{flchrt}) and an in-house code named Completely Adaptive Random Mutation Hill Climbing (CARMHC: right side in Fig.\ref{flchrt}), which was previously developed for other optimization problems.\cite{Sarkar2010, Sarkar2012, Sarkar2013, Sarkar2013a} The GA starts with a population of potential solutions $S=\lbrace s \rbrace _i$, $i=1$, $n_p$, where $n_p$ is the cardinality of the population, that is allowed to undergo a simulated evolution in a sense that in each generation (iteration step) the relatively good solutions are allowed to stay on and reproduce while the bad ones die out, moving the population toward better solutions. The evolution usually starts with a randomly generated population of individuals (candidate solutions). Since the current optimization problem involves very costly {\em{ab initio}} calculations, random starting points may require unnecessarily large number of generations and hence longer time to home into the important region of the search space. Therefore, we have employed CARMHC together with the atomic PAW data-set generator program ATOMPAW to quickly generate $2 \times n_p$ individuals for the starting GA population. The process, repeated $2 \times n_p$ times, starts with a typical string, e.g., a JTH parameter set, changes $r_c$ randomly within a constrained range of values, and maximizes $f^s(O^s_{\rm atom})$. Then, the GA augmented with ATOMPAW and Quantum ESPRESSO codes maximizes $f^s(O^s_{\rm solid})$ to optimize the PAW data-sets for an arbitrary volume ranges. During optimization, the GA also monitors $f^s(O^s_{\rm atom})$ so that its value remains larger than $f^s(O^s_{\rm atom})$ of the initial JTH data-set.

For each atom in our study, we have used the corresponding elemental solid structure used by Lejaeghere {\em{et al.}}\cite{Lejaeghere2014,Lejaeghereaad3000} in the Delta package.\cite{Deltav3.1} The total energies for each solid are evaluated at 15 values of $V^{1/3}$ uniformly spaced such that $0.78 V_0^{1/3} \le V^{1/3} \le 1.06 V_0^{1/3}$. The chosen range of volumes allows us to probe compressed volumes as small as $V_0/2$. We have tested different schemes to generate pseudized partial waves, projectors, and local pseudopotential and report only the best results (see Supplemental Material\cite{SM}).

\section{Methodology}   \label{ethodology}
\begin{figure*}[bth!] 
\includegraphics[width=0.75\linewidth]{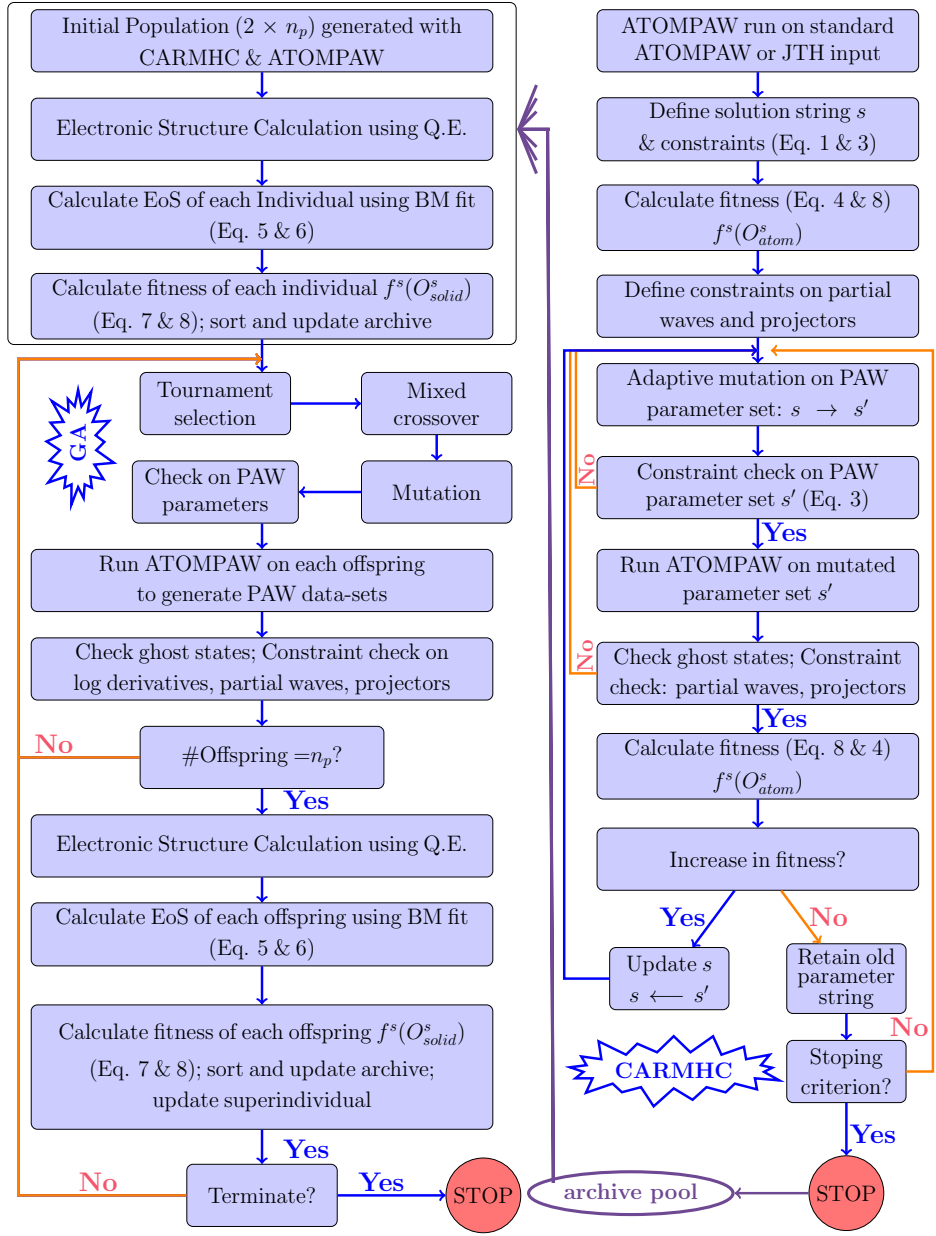} 
\caption{Flowchart for the whole procedure of generating optimized PAW data-sets. The right part represents a combination of ATOMPAW and CARMHC algorithms and quickly generates an initial population for the GA to start with. The left part documents hybridization of the GA with ATOMPAW and Quantum Espresso distributions. The GA optimizes the free parameter set which is used by ATOMPAW and Quantum ESPRESSO to generate the optimum PAW data-set.}\label{flchrt} 
\end{figure*}

We have devised a sequence of steps (see flowchart in Fig. \ref{flchrt}) based on a variety of algorithms inspired by genetics to find the string $s^{\star}$ specifying the PAW dataset that optimally reproduces the high pressure behavior of solids. The process starts with the following initial information for a given atom:
\begin{enumerate}
\item Choose a pseudopotential generation scheme and PAW parameters to use for an atom. The choice includes the number of variables $M$ (Eq. \ref{stringdef}) and their ranges of values as specified in Eq. \ref{range}.
\item Choose an initial parameter set $s$. This can be taken from an existent PAW data-set from various sources ({\em{e.g.,}}  JTH\cite{Jollet2014} or ATOMPAW\cite{holzwarth:2001,Torrent20101862}) or can  be generated as a random guess.
\item Determine the all-electron equation of state parameters $V^{AE}_0$,  $B^{AE}_0$, and $B'^{AE}_0$ for the structure chosen to represent the atom in the solid state environment. These are needed in order to evaluate Eq. \ref{obj}. In this work, the full-potential linearized augmented wave code package WIEN2k \cite{wien2k:2001} is used.
\end{enumerate}

\begin{figure}[bth!] 
\includegraphics[width=0.75\linewidth]{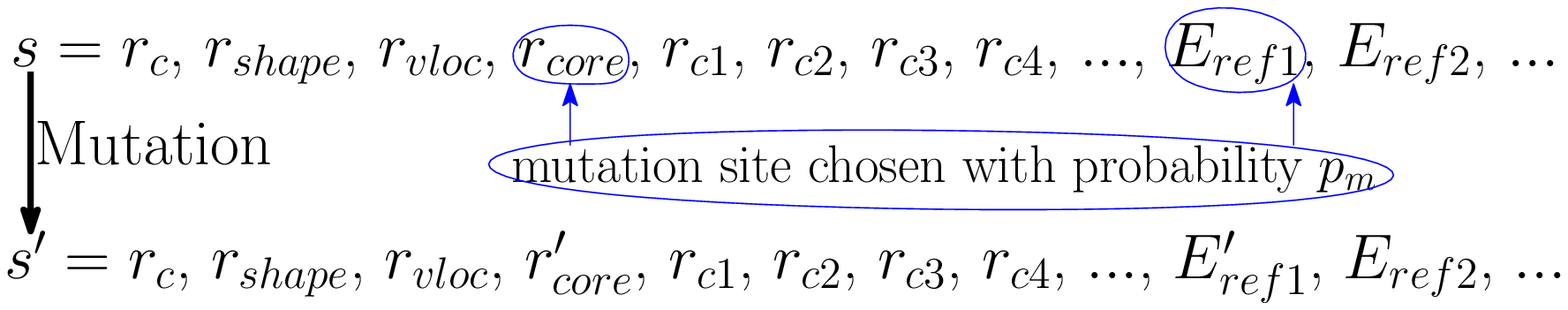} 
\caption{Mutations or small perturbations to the string $s$ during the course of minimizing $O_{atom}$ with mutation intensity $\Delta_m$ in a continuous space. One or more elements in $s$ have been chosen for mutation with probability $p_m$. Both parameters $p_m$ and $\Delta_m$ in the meta-heuristics CARMHC have been adaptively determined based on previous exploratory performances of this algorithm.\cite{Sarkar2010, Sarkar2013a}}\label{mutn} 
\end{figure}

The CARMHC algorithm starts with one standard PAW atomic data-set $s$ as initial guess and employs mutation, as the only evolutionary process for minimizing the area between the logarithmic derivative curves generated by all electron and PAW calculations. Mutations may be visualized as little perturbations to $s$ by noise (a function of mutation intensity, $\Delta_m(v_j)$, with a mutation probability $p_m$ in a continuous space. Fig. \ref{mutn} illustrates the process. Mutation in CARMHC involves two flexible parameters including the mutation probability ($p_m$) and the mutation intensity ($\Delta_m$), both of which have been dynamically adjusted according to previous experience with this algorithm. For more details about CARMHC algorithm, we refer to the existing literature.\cite{Sarkar2013a} The underlying principle is to randomly search for a better solution ($s^{\prime}$) in the neighborhood of the current solution ($s$) by mutating the string $s$ with some chosen meta-heuristics (problem-independent rule to perturb the solution string). The atomic program generates the mutated PAW data-set in every generation and the atomic fitness value $f^s_{\rm atom}$ is evaluated according to Eqs. (\ref{logd}) and (\ref{eqfit}). If the mutation results in a data-set $s^{\prime}$ with higher fitness compared to the current one, $s$, the new solution is stored as $s$, otherwise, the current solution, $s$, is retained. The process  continues until no mutation causes any increase in the fitness of the current solution for a certain number of generations ($\sim 50$). The solution string at this stage is returned as the result. $2 \times n_p$ such runs of the CARMHC code produces $2 \times n_p$ individuals or solution strings having different $r_c$ values for a given scheme and stored in an external archive. Thus CARMHC is a smartly coded prescription to quickly generate initial trial solutions for the GA. More details can be found in the cited references.\cite{Sarkar2010, Sarkar2012, Sarkar2013, Sarkar2013a}

Once the external archive is populated with $2 \times n_p$ solution strings (or individuals, or chromosomes), they are fed into a second in-house GA code's initial population and are evaluated for their fitness (Eqs. \ref{obj} and \ref{eqfit}). The general features of the GA  are shown on the left side of Fig. \ref{flchrt}. This archive keeps the best ever solutions found in the course of GA runs, {\em{i.e.}}, the archive is updated in each generation ($t$) by replacing the dominated solutions with more fit individuals from the current generation. The $n-$member Tournament selection procedure\cite{DavidE.Goldberg} is then used to prepare a mating pool with population size $n_p$ from the archive having $2 \times n_p$ strings. Strings from the archive with above-average fitness have greater chance to enter the mating pool. A tournament is held among $n$ randomly picked competitors from the archive. The individual with the best fitness value among those random $n$ tournament competitors (winner of the tournament) is then copied into the mating pool. The genetic operator crossover is then applied with probability $p_c$ to randomly selected pairs of solution strings from the mating pool. $p_c$ is the probability of using crossover for creating offsprings. It should be noted here that the crossover operator was designed to share information between two individuals (randomly chosen parents) by swapping or intermingling their elements (genetic materials), with the possibility that good chromosomes may generate promising descendants. The operator is applied to randomly selected pairs of individuals until adequate numbers of offsprings are produced. We use two types of crossover operators: 1) for pairs of individuals with different fitness values we choose arithmetic crossover\cite{Michalewicz1996}; 2) for pairs with similar fitness values we choose BLX$-\alpha$ crossover.\cite{ESel} The arithmetic crossover operates between two randomly selected individuals ($s_i$ and $s_j$) from the mating pool to produce two descendants $o_1$ and $o_2$. This type of crossover helps to speed up the algorithm convergence. Once the individuals are chosen to undergo crossover, the next step is to select the crossover site(s). In single point arithmetic crossover, one crossover site ($t$) is randomly selected from [$1,2, ..., M-1$], where $M$ is the number of variables of an individual. The two randomly chosen parents from the mating pool, $s_i$ and $s_j$, are intermingled beyond that cross site $t$ by complementary linear combinations (convex combination) of $s_i$ and $s_j$ using an arithmetic mean. The $k^{th}$ entry of two offsprings are determined as follows:
\begin{eqnarray}
o_{1}^{k}&=&\begin{cases} \alpha \times s_{i}^{k}+(1-\alpha )\times s_{j}^{k}, \text{ if } k\ge t \\ s_{i}^{k},\text{if }  k<t  \end{cases}\\
o_{2}^{k}&=&\begin{cases} (1- \alpha )\times s_{i}^{k}+\alpha\times s_{j}^{k}, \text{ if } k\ge t \\ s_{j}^{k},\text{if }  k<t\end{cases}
\end{eqnarray}
where $\alpha >0 $ is constant. Here, we have used two-point arithmetic crossover, where three crossover points ($t_1,t_2,t_3$) are chosen at random with no duplicates and sorted into ascending order. The genetic materials of the two parent between the first two points ($t_1 \: \& \:t_2$) and beyond the third cross site ($t_3$) are combined by taking the weighted sum.
\begin{eqnarray}
o_{1}^{k}=\begin{cases} \alpha \times s_{i}^{k}+(1-\alpha )\times s_{j}^{k},\text{ if } t_1\le k \le t_2 \text{ or } k\ge t_{3} \\ s_{i}^{k}, \text{ otherwise } \end{cases}\\   
o_{2}^{k}=\begin{cases}  (1- \alpha ) \times s_{i}^{k}+\alpha\times s_{j}^{k},\text{ if } t_1\le k \le t_2 \text{ or } k\ge t_{3}  \\ s_{j}^{k}, \text{ otherwise }   \end{cases}
\end{eqnarray}

BLX$-\alpha$ crossover expands the range of arithmetic crossover. It generates a single offspring by blending two randomly selected floating point parent vectors, $s_i$ and $s_j$, from the mating pool. The $k^{th}$ entry of an offspring is determined as follows: 
\begin{eqnarray}
&&o_{k} = {\rm R} \left( \frac{}{}(L_k -\alpha . I_k), \; (U_k +\alpha . I_k) \frac{}{} \right)  \\
&&{\rm where}\;\; U_k = max\left(s_{i}^{k}, s_{j}^{k} \right), \;\; L_k = min\left(s_{i}^{k}, s_{j}^{k} \right),\nonumber\\
&&{\rm and}\;\; I_k=  U_k - L_k \nonumber 
\end{eqnarray}
${\rm R}(a,b)$	is a uniform random number between $a$ and $b$. The user-defined parameter $\alpha$ is usually set to 0.5. BLX$-\alpha$ crossover is applied to maintain the population diversity since there is a good chance that solution strings having similar fitness value also have similar genetic materials.

The new population of $n_p$ offsprings is allowed to undergo mutation (Fig. \ref{mutn}) with probability $p_m$ on each string element. $p_m$ is the probability of modifying one or more elements of an individual. The new individuals are then subjected to constraints satisfaction. If any variable of the individuals exceeds its predefined maximum (or goes below the minimum), it is truncated to that maximum value (or minimum value). The ATOMPAW program then generates $n_p$ different data-sets. Corresponding constraint checks on logarithmic derivatives and wave functions are carried out, such as:
\begin{enumerate}
\item $f^s(O^s_{\rm atom})$ of each generated data-set should not be smaller than that of the initial guess {\em{i.e.,}} $f^s(O^s_{\rm atom})$ of JTH or ATOMPAW data-set set as standard. 
\item The logarithmic energy derivative of the radial wave functions for the exact atomic problem and the pseudized problem should superimpose as much as possible. There should be no discontinuity in the logarithmic derivative curve in the range of $-4.0 \leq  E_0 \leq 4$ Rydberg. 
\item Partial and pseudized partial-waves should meet near or after the last maximum (or minimum).
\item Partial-waves, pseudized partial-waves and projector functions should have the same order of magnitude to avoid numerical instability and to promote good transferability.\cite{Jollet2014}
\end{enumerate} 

The entire process along with execution of ATOMPAW code is repeated until an adequate number ($n_p$) of PAW data-sets are produced. These $n_p$ different data-sets are then utilized for {\em{ab initio}} calculations with the Quantum ESPRESSO. The new individuals (offsprings) are evaluated for their fitness (Eqs. \ref{obj} and \ref{eqfit}). Solution strings from parents and offsprings form the new mating pool for the next generation through Tournament selection (on $2 \times n_p$ individuals). Thus the evolution of the individuals with selection, crossover, and mutation ensure that progressively better and better solutions are discovered as generations elapse. The process continues until the fitness stops evolving. After a number of such cycles are repeated, usually the average fitness ($f_{av}$) and the maximum fitness ($f_{max}$) of the population saturates. The string with maximum fitness (super individual) is then hopefully the solution we are looking for. The fitness (Eqs. \ref{eqfit} and \ref{obj}) of the candidate solutions in the population quantifies the difference between WIEN2k (AE-FLAPW) and PAW results. The flowchart of the complete algorithm is shown in Fig. \ref{flchrt}. The technique is a non-deterministic evolutionary search procedure augmented with deterministic bias from ATOMPAW and Quantum ESPERSSO results. Our goal here is to introduce a hybrid optimization technique to generate PAW data-sets with uniform performance up to or at specific high pressure regions for selected elemental crystal structures from a benchmark set.\cite{Lejaeghere2014} Therefore the present hybrid algorithm is essentially a goal-directed random search procedure in which the target is set to the WIEN2k equation of state. The target can be given by any other AE-FLAPW implementation.

\section{Computational Details   \label{Comp}}
The reference all-electron calculations are carried out using the all-electron full-potential linearized augmented-plane-wave approach,\cite{Yu91} as implemented in the WIEN2k code.\cite{wien2k:2001} In this method, the wavefunctions are expanded in terms of spherical harmonics inside the muffin-tin spheres of radius $R_{MT}$ surrounding each atom and in terms of simple plane waves in the interstitial region. In order to treat the high-pressure behavior of solids, $R_{MT}$ values are reduced by 25$\%$ from their default values listed in Table \ref{table:radius}. In order to ensure the accuracy of all-electron results we chose to use large convergence parameters, {\em{e.g.}}, the cut-off wave vector of plane wave expansion in the interstitial region is set to $K_{max}=10.0/R_{MT}$, which is 43$\%$ larger than the default and well-converged value. The Brillouin zone integrations used the same grid in WIEN2k and PAW calculations.  
\begin{table}
\caption{\label{table:radius} 
Reference muffin-tin radii $R_{MT}$ (in Bohr units) used in AE-FLAPW calculations in this study.} 
\begin{tabular}{|c|ccccc|} \hline
            & C    &  Mg    & Al     & Si   &     Fe  \\ \hline
$R_{MT}$    & 1.03 &  1.20   & 1.80    & 1.60 &  1.40  \\ \hline
\end{tabular}
\end{table}

All-electron and PAW-Quantum ESPRESSO\cite{PWscf:2009} calculations are performed using the Perdew-Burke-Ernzenhof (PBE) functional.\cite{Perdew1996} In order to focus on the accuracy of PAW calculations without regard for efficiency, we set a very high convergence criterion: a plane-wave expansion of $|{\bf{k}}+{\bf{G}}|^2 \le 100$~Ry to represent the wave-function and $|{\bf{G}}|^2 \le 500$~Ry to represent the density. A Fermi-Dirac or Gaussian smearing function with width of 0.001 Ry is used for both data-set and AE calculations. The numerical settings for the evolutionary algorithms are given in Table \ref{table:EC}. Total energies are evaluated using Quantum ESPRESSO with GBRV, JTH, HGH and EPAW for 15 equidistant points between $0.78$ $a_0$ and $1.19$ $a_0$, where $a_0$ is the equilibrium lattice constant. Exactly the same lattice parameters are used for VASP and WIEN2k calculations. Pressures have been calculated using a third order Birch-Murnaghan fit.\cite{Murnaghan1944, Birch1947}
\begin{table}
\caption{\label{table:EC} 
Parameters used in the evolutionary algorithms in this study.} 
\begin{scriptsize}
\begin{tabular}{|c|llll|} \hline
Method	&Parameters	             		&Starting       	&Maximum 	&Minimum       \\ \hline
CARMHC	&Mutation probability ($p_m$)		& 0.33	        &    0.33	&  0.05        \\
	&Mutation intensity ($\Delta_m$)	    & 0.1	        &    0.2	&  1.0E-10     \\\hline
GA	& $\#$ Population ($n_p$)		        & 10         	&    -- 	&  --         \\
	&Crossover probability ($p_c$)		    & 0.7 -- 0.8	&    0.9	&  0.1         \\
	&Crossover intensity ($\alpha$)	    & 0.6	        &    0.8	&  0.1         \\
	&$\alpha$ in BLX$-\alpha$ crossover	& 0.6	        &    --	    &  --          \\
	&Mutation probability ($p_m$)		    & 0.1	        &    0.33	&  0.05        \\
	&Mutation Intensity ($\Delta_m$)	    & 0.01	        &    0.2	&  1.0E-10     \\
	&No. of members in                 	&  	            &         	&              \\ 
	&tournament selection	                & 2	            &     --	&  --           \\ \hline	
\end{tabular}
\end{scriptsize}
\end{table}

The constraints on the radial parameters in the solution string $s$ are imposed such that the augmentation radius $r_c$ is the largest of all of the matching radii including $r_{shape}$, $r_{vloc}$, $r_{core}$, and $r_{ci}$. $r_{c}$ should be large enough to make the PAW potential as soft as possible.\cite{holzwarth:2001,Tackett2001} Again, it is very much essential to have very small augmentation regions ($r_{c}$), without resulting in sphere overlap, and inclusion of semi-core electrons in the valence to generate PAW data-sets for materials simulations at high pressures. In general, the energy range for evaluating the logarithmic derivatives in Eq. (\ref{logd}) is taken to be $0 \le E \le 4$ Ry, but a check is necessary in the range of negative energies to make sure that there are no ghost states. The constraints are discussed in section \ref{ethodology} in more detail. 

\section{Results and Discussion \label{RD}}

To assess the value of the current approach, we compare the performance of ``EPAW" data-sets (Evolutionarily optimized PAW data-sets) with those of GBRV ultra-soft pseudopotentials,\cite{Garrity2014} HGH norm-conserving data-sets,\cite{Hamann1979, Vanderbilt1985} the recently released JTH PAW data-sets,\cite{Jollet2014} and VASP (PAW)\cite{Kresse1999}. This is done by monitoring the difference between generated EoSs and the all-electron EoS, i.e., $\vert \Delta P \vert$ against $V$ (Eq. \ref{obj}) predicted by different schemes over an enlarged pressure range. 

The first test of this approach concerned a non-magnetic bcc elemental crystal of iron. The EPAW algorithm used the JTH PAW data-set parameters as the initial string. Fig. \ref{FeEOS}, clearly reveals that EPAW data-set outplays all the others, in terms of both $\vert \Delta P (V)\vert$ and performance uniformity throughout the entire range of volume compression. Except for JTH, all the other atomic data-sets perform well around zero pressure, but this picture changes at high pressures. The higher the pressure, the higher the deviation from WIEN2k results. Although GBRV data-set performs well and uniformly, the optimized PAW data-set (EPAW) is a little more accurate. Therefore the proposed approach successfully improves the quality of the PAW data-sets for bcc nonmagnetic iron. The important computational settings for the calculations performed and equation-of-state parameters have been included in Table \ref{tab:table1}. While the optimization calculations are all performed using the Quantum ESPRESSO code,\cite{PWscf:2009} we examined also the performance of the optimal data-sets using the ABINIT code.\cite{abinit:2009} In principle, these two independent codes have the same formalism implemented. By comparing their performances with the same PAW data-sets we are able to assess numerical errors related with method implementation.

\begin{table*}
\caption{\label{tab:table1}  Computational settings and predicted equation-of-state parameters of some selected elemental crystals from the benchmark set by Lejaeghere \emph{et al}. \citep{Lejaeghere2014} are listed for each method of data-set for comparison.  }
\begin{ruledtabular}
\begin{scriptsize}
\begin{tabular}{|ccccccc|}
        & Code    &          $k-$point grid       & valence          &$V_0[\AA ^3/atom]$&$B_0[GPa]$&$B^{\prime}[-]$  \\ \hline
C & WIEN2k(target)& $18 \times 18 \times 4 $ & $2s\:2p$         & $11.647 (\pm 0.002)$ & $207.363 (\pm 0.359)$ & $3.576  (\pm 0.003)$   \\        
        & GBRV    & $18 \times 18 \times 4 $ & $2s\:2p$         & $11.635 (\pm 0.003)$ & $205.950 (\pm 0.437)$ & $3.598  (\pm 0.004)$   \\        
        & HGH     & $18 \times 18 \times 4 $ & $2s\:2p$         & $11.644 (\pm 0.003)$ & $206.270 (\pm 0.470)$ & $3.547  (\pm 0.004)$   \\        
        & JTH     & $18 \times 18 \times 4 $ & $2s\:2p$         & $11.642 (\pm 0.002)$ & $209.294 (\pm 0.273)$ & $3.644  (\pm 0.002)$   \\        
        & VASP    & $18 \times 18 \times 4 $ & $2s\:2p$         & $11.637 (\pm 0.002)$ & $207.728 (\pm 0.368)$ & $3.572  (\pm 0.003)$   \\        
        & EPAW(QE)& $18 \times 18 \times 4 $ & $2s\:2p$         & $11.654 (\pm 0.003)$ & $206.839 (\pm 0.396)$ & $3.577  (\pm 0.003)$   \\
    & EPAW(ABINIT)& $18 \times 18 \times 4 $ & $2s\:2p$         & $11.664 (\pm 0.003)$ & $206.671 (\pm 0.400)$ & $3.578  (\pm 0.003)$   \\   \hline
Mg& WIEN2k(target)& $14 \times 14 \times 8 $ & $2s\:2p\:3s$     & $23.601 (\pm 0.007)$ & $33.227  (\pm 0.097)$  & $3.899 (\pm 0.006)$   \\
        & GBRV    & $14 \times 14 \times 8 $ & $2s\:2p\:3s$     & $22.942 (\pm 0.005)$ & $36.372  (\pm 0.086)$  & $3.866 (\pm 0.005)$   \\              
        & HGH     & $14 \times 14 \times 8 $ & $3s$             & $23.286 (\pm 0.005)$ & $35.040  (\pm 0.076)$  & $3.796 (\pm 0.004)$   \\              
        & JTH     & $14 \times 14 \times 8 $ & $2s\:2p\:3s$     & $23.341 (\pm 0.007)$ & $36.706  (\pm 0.111)$  & $3.850 (\pm 0.006)$   \\             
        & VASP    & $14 \times 14 \times 8 $ & $2s\:2p\:3s$     & $22.952 (\pm 0.004)$ & $36.459  (\pm 0.070)$  & $3.859 (\pm 0.004)$   \\              
        & EPAW(QE)& $14 \times 14 \times 8 $ & $2s\:2p\:3s$     & $23.192 (\pm 0.005)$ & $36.316  (\pm 0.070)$  & $3.838 (\pm 0.004)$   \\
    & EPAW(ABINIT)& $14 \times 14 \times 8 $ & $2s\:2p\:3s$     & $22.946 (\pm 0.006)$ & $36.480  (\pm 0.098)$  & $3.865 (\pm 0.006)$   \\         \hline
Al& WIEN2k(target)& $36 \times 36 \times 36$ & $2s\:2p\:3s\:3p$ & $16.492 (\pm 0.020)$ & $82.243  (\pm 0.919)$  & $4.015 (\pm 0.022)$   \\
        & GBRV    & $36 \times 36 \times 36$ & $3s\:3p$         & $16.501 (\pm 0.020)$ & $81.862  (\pm 0.922)$  & $4.027 (\pm 0.022)$    \\        
        & HGH     & $36 \times 36 \times 36$ & $3s\:3p$         & $16.473 (\pm 0.020)$ & $81.509  (\pm 0.883)$  & $3.949 (\pm 0.021)$    \\        
        & JTH     & $36 \times 36 \times 36$ & $3s\:3p$         & $16.477 (\pm 0.021)$ & $82.832  (\pm 0.957)$  & $4.047 (\pm 0.023)$    \\        
        & VASP    & $36 \times 36 \times 36$ & $3s\:3p$         & $16.479 (\pm 0.021)$ & $81.861  (\pm 0.953)$  & $4.036 (\pm 0.023)$    \\        
        & EPAW(QE)& $36 \times 36 \times 36$ & $2s\:2p\:3s\:3p$ & $16.502 (\pm 0.021)$ & $82.252  (\pm 0.936)$  & $4.009 (\pm 0.022)$    \\ 
    & EPAW(ABINIT)& $36 \times 36 \times 36$ & $2s\:2p\:3s\:3p$ & $16.488 (\pm 0.022)$ & $82.422  (\pm 1.020)$  & $4.014 (\pm 0.024)$    \\ \hline
Si& WIEN2k(target)& $12 \times 12 \times 12$ & $2p\:3s\:3p$     & $20.476 (\pm 0.018)$ & $93.291  (\pm 0.749)$  & $3.780 (\pm 0.014)$   \\        
        & GBRV    & $12 \times 12 \times 12$ & $3s\:3p$         & $20.451 (\pm 0.020)$ & $92.855  (\pm 0.797)$  & $3.782 (\pm 0.016)$   \\             
        & HGH     & $12 \times 12 \times 12$ & $3s\:3p$         & $20.377 (\pm 0.019)$ & $92.162  (\pm 0.757)$  & $3.755 (\pm 0.015)$   \\             
        & JTH     & $12 \times 12 \times 12$ & $3s\:3p$         & $20.456 (\pm 0.020)$ & $93.131  (\pm 0.817)$  & $3.788 (\pm 0.016)$   \\             
        & VASP    & $12 \times 12 \times 12$ & $3s\:3p$         & $20.495 (\pm 0.018)$ & $92.260  (\pm 0.733)$  & $3.786 (\pm 0.014)$   \\             
        & EPAW(QE)& $12 \times 12 \times 12$ & $3s\:3p$         & $20.457 (\pm 0.019)$ & $93.689  (\pm 0.791)$  & $3.778 (\pm 0.015)$   \\ 
    & EPAW(ABINIT)& $12 \times 12 \times 12$ & $3s\:3p$         & $20.467 (\pm 0.019)$ & $93.475  (\pm 0.776)$  & $3.782 (\pm 0.015)$   \\ \hline    
Fe      & WIEN2k  & $16 \times 16 \times 16$ & $3s\:3p\:3d\:4s$ & $10.566 (\pm 0.012)$ & $266.941 (\pm 3.516)$ & $4.315 (\pm 0.039)$    \\             
        & GBRV    & $16 \times 16 \times 16$ & $3s\:3p\:3d\:4s$ & $10.505 (\pm 0.006)$ & $273.253 (\pm 1.835)$ & $4.304 (\pm 0.020)$    \\        
        & HGH     & $16 \times 16 \times 16$ & $3s\:3p\:3d\:4s$ & $10.617 (\pm 0.006)$ & $270.300 (\pm 1.659)$ & $4.299 (\pm 0.018)$    \\        
        & JTH     & $16 \times 16 \times 16$ & $3s\:3p\:3d\:4s$ & $10.125 (\pm 0.129)$ & $364.179 (\pm 52.30)$ & $4.149 (\pm 0.511)$    \\        
        & VASP    & $16 \times 16 \times 16$ & $3s\:3p\:3d\:4s$ & $10.678 (\pm 0.003)$ & $259.766 (\pm 0.921)$ & $4.275 (\pm 0.010)$    \\        
        & EPAW(QE)& $16 \times 16 \times 16$ & $3s\:3p\:3d\:4s$ & $10.576 (\pm 0.007)$ & $265.191 (\pm 1.865)$ & $4.322 (\pm 0.021)$    \\ 
    & EPAW(ABINIT)& $16 \times 16 \times 16$ & $3s\:3p\:3d\:4s$ & $10.574 (\pm 0.007)$ & $265.574 (\pm 1.874)$ & $4.325 (\pm 0.021)$    \\ \hline    
\end{tabular}
\end{scriptsize}
\end{ruledtabular}
\end{table*}

\begin{figure}[bth!]
\includegraphics[width=0.75\linewidth]{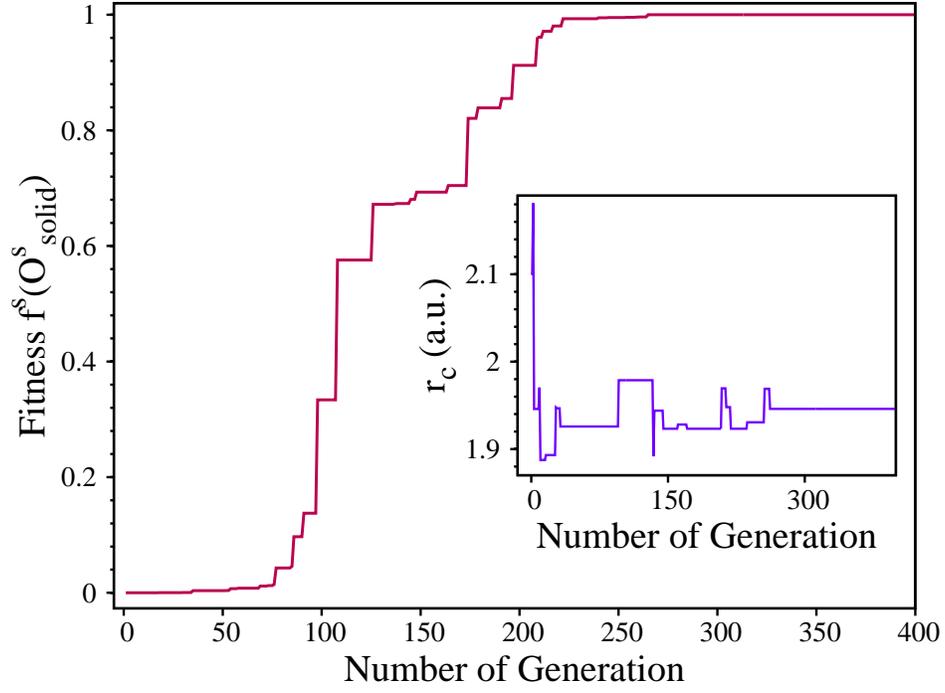}
\caption{Fitness evolution profile of the best evolving PAW data-set for bcc elemental iron (nonmagnetic) crystal. The inset figure displays the way $r_c$ changes with GA generation during the course of PAW parameters optimization.}\label{ABSFTNS}
\end{figure}

The workability of the GA-based strategy for PAW parameters optimization is illustrated in Fig. \ref{ABSFTNS} showing the evolution of the fitness function of the best evolving string of bcc (nonmagnetic) iron. The raw fitness of the best evolving string in the population has been scaled from 0 to 1. The corresponding evolution of $r_c$ is displayed in the inset. We note that the proposed hybrid soft-computing method takes $\sim 220$ generations ($\sim 2000$ EoS calculations) for iron to move into the global minimum region of the search space. For the other elements it takes on average $\sim 100$ generations ($\sim 1000$ EoS calculations). The convergence speed can be improved by fine tuning the genetic parameters shown in table \ref{table:EC}. The sharp rise of the fitness value in the initial evolution region is mainly controlled by crossover operations, while the final flat portion of the evolution profile is dominated by mutation.

To inquire about the transferability of the generated EPAW data-set to a greater extent, we first optimize the PAW data-set for carbon in the graphite structure and test its performance in the diamond structure. Figure (\ref{CEOS}) clearly indicates that in both the cases, the optimized PAW data-set (EPAW) performs more uniformly and produces pressures closer in overall to the WIEN2k pressures.

In the previous examples of carbon and iron, the weight factors ($\omega _i$) in Eq. \ref{obj} are assigned to 1. To understand the effect of the $\omega _i$ in Eq. \ref{obj} we optimize some PAW data-sets with $\omega _i$ increasing linearly from $\omega _{min}$ at 1.06 $a_0$ to $\omega _{max}$ at 0.78 $a_0$. In hcp Mg $\omega _{min}$ and $\omega _{max}$ are 0.9 and 1.0, while for fcc Al and diamond Si the lower and upper bounds to $\omega _i$ are 0.95 and 1.0 respectively. As expected, the EPAW data-set performs better at higher pressures (Figs. \ref{MgEOS}, \ref{SiEOS} and \ref{AlEOS}). JTH data-sets are taken as the initial guesses in these cases. Since results using JTH data-sets deviate more from the target at high pressures, we increase the weight on that region for optimizing the set of PAW parameters. This also helps the algorithm to quickly home into the important region of the search space. The overall performance of the resulting EPAW data-sets are improved with respect to the other atomic data-sets. The EPAW data-set for Si shows comparable performance with GBRV, JTH, and VASP data-sets around equilibrium volume and produces smaller pressure differences from WIEN2k results in the high-pressure region. In the case of aluminum (Fig. \ref{AlEOS}), GBRV, VASP, and EPAW perform almost equally well compared to the other two data-sets.

\section{Goodness measures for atomic data-sets} \label{GDP}
Recently Lejaeghere \emph{et al}.\cite{Lejaeghere2014, Lejaeghereaad3000} proposed a numerical measure, $\Delta$, to quantitatively assess the quality of a DFT potential. $\Delta$ is a measure of the distance between energy-volume curves produced by the AE (WIEN2k or any other AE-FLAPW code) and data-set ($X$) calculations.
\begin{eqnarray}
\Delta (\text{AE,X})= \sqrt{\int_{V_{1}}^{V_{2}} \frac{\left(E_{\text{X}}(V)-E_{\text{AE}}(V)\right)^2}{V_{2} -V_{1} }dV}
\end{eqnarray}
Here $E(V)$ represents the energy per atom at volume $V$ and the inter-code energy difference is to be integrated between the volume $V_{1}$ and $V_{2}$. Jollet \emph{et al}.\cite{Jollet2014} further renormalized the $\Delta$ gauge by a factor involving the zero pressure equilibrium volume ($V_{0}$) and a bulk modulus ($B_{0}$).
\begin{eqnarray}
\Delta_{1}(\text{AE,X}) = \frac{V_{0}^{\text{X}}B_{0}^{\text{X}}}{V_{0}^{\text{AE}}B^{\text{AE}}}\Delta(\text{AE,X}),
\end{eqnarray}
Both of $\Delta$ and $\Delta_{1}$ have been implemented in the $\Delta$ calculation package 3.0\cite{Deltav3.1} along with another equivalent prescription \cite{Lejaeghereaad3000} ($\Delta_{rel}$):
\begin{eqnarray}
\Delta_{rel}(\text{AE,X}) = 2 \sqrt{\frac{\int_{V_{1}}^{V_{2}} \left(E_{\text{X}}(V)-E_{\text{AE}}(V)\right)^2 dV}{\int_{V_{1}}^{V_{2}} \left(E_{\text{X}}(V)+E_{\text{AE}}(V)\right)^2 dV}}\end{eqnarray}
All formulations above are based on the root mean square of the area between the AE and atomic data-set energy-volume curves. These measures capture the degree of similarity between results obtained with two different methods. Presently available atomic data-set libraries are tuned in order to reproduce AE-FLAPW results around the zero pressure equilibrium lattice constant(s) and hence they are generated for a small pressure range. The atomic data-sets from standard libraries\cite{Garrity2014,Jollet2014,dal_corso_pseudopotentials_2014} usually behave in a uniform way around zero pressure. Consideration of an enlarged pressure range introduces large deviations and fluctuating behavior in the performance of these standard data-sets.

\begin{figure}[bth!]
\includegraphics[width=0.95\linewidth]{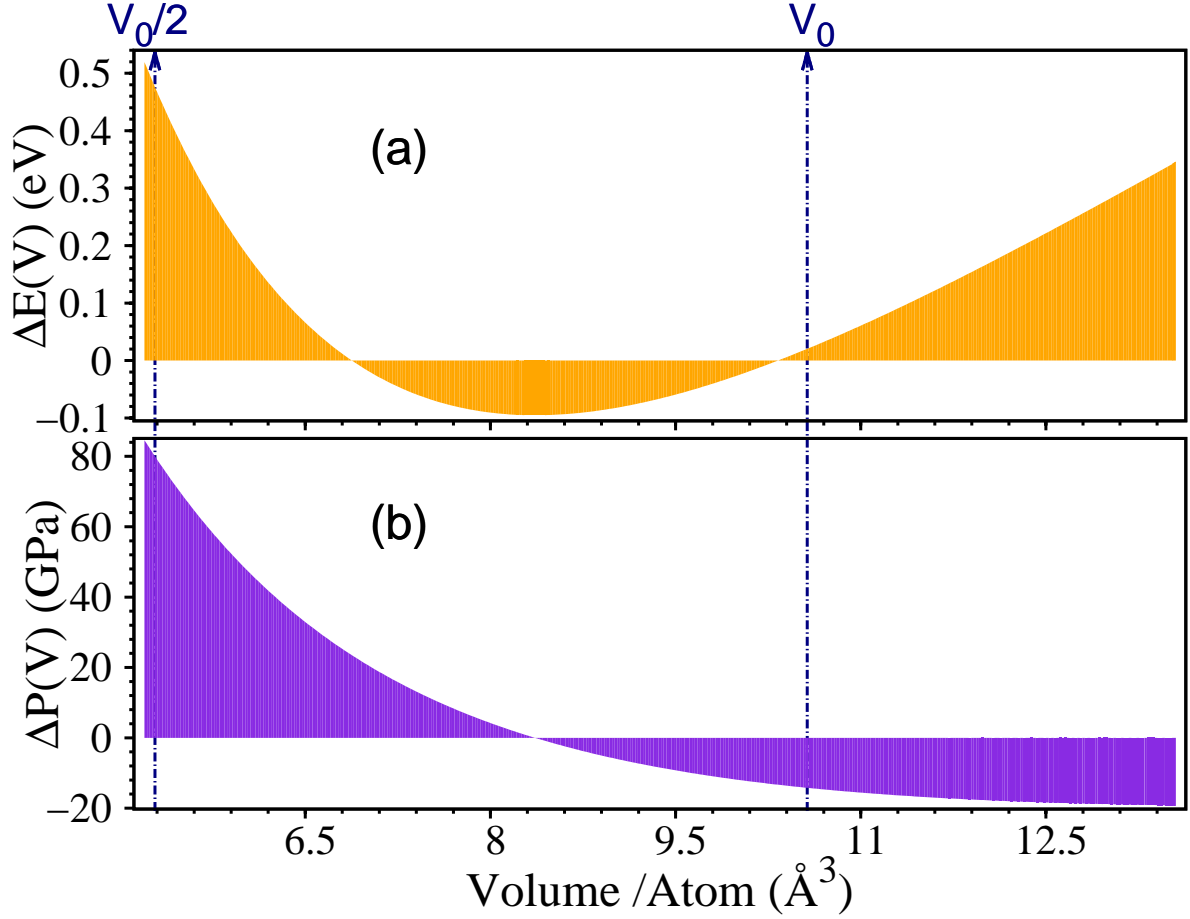}
\caption{(a) Energy difference and (b) pressure difference for bcc (nonmagnetic) iron between AE-FLAPW calculations (WIEN2k) and a JTH PAW calculations (Quantum ESPRESSO). We aimed at minimizing the colored areas between $0.475$ $V_0$ and $1.19$ $V_0$, where $V_0$ is the equilibrium volume. In some cases the pressure differences are subjected to a volume dependent weight in the PAW optimization process (see text).}\label{wflw}
\end{figure}

In order to extend these ideas and consider the behavior of data-sets under high-pressures, we are motivated to define new ``goodness" measures. Fig. \ref{wflw}(a) shows the total energy difference between JTH-PAW and AE-FLAPW results for iron (Fe) for a large volume range, where $V_0$ is the AE zero pressure volume. $\Delta E$ is small around $V_0$. However, $\Delta E$ can be as large as 500 meV when the volume is compressed to (${V_0}/{2}$). Similarly, the pressure difference can also be used to describe the difference between AE-FLAPW and PAW results. Fig. \ref{wflw}(b) shows the pressure difference ($\Delta P$) between JTH-PAW and AE-FLAPW results. Using Perdew-Burke-Ernzenhof (PBE) functional\cite{Perdew1996}, the AE-FLAPW pressure of bcc-Fe at (${V_0}/{2}$) is around 800 GPa, while the JTH-PAW pressure is $\approx$ 80 GPa (10$\%$) larger for the same volume. For most atomic data-sets, deviations from AE-FLAPW pressure increase with compression. 

\begin{figure}[bth!]
\includegraphics[width=0.75\linewidth]{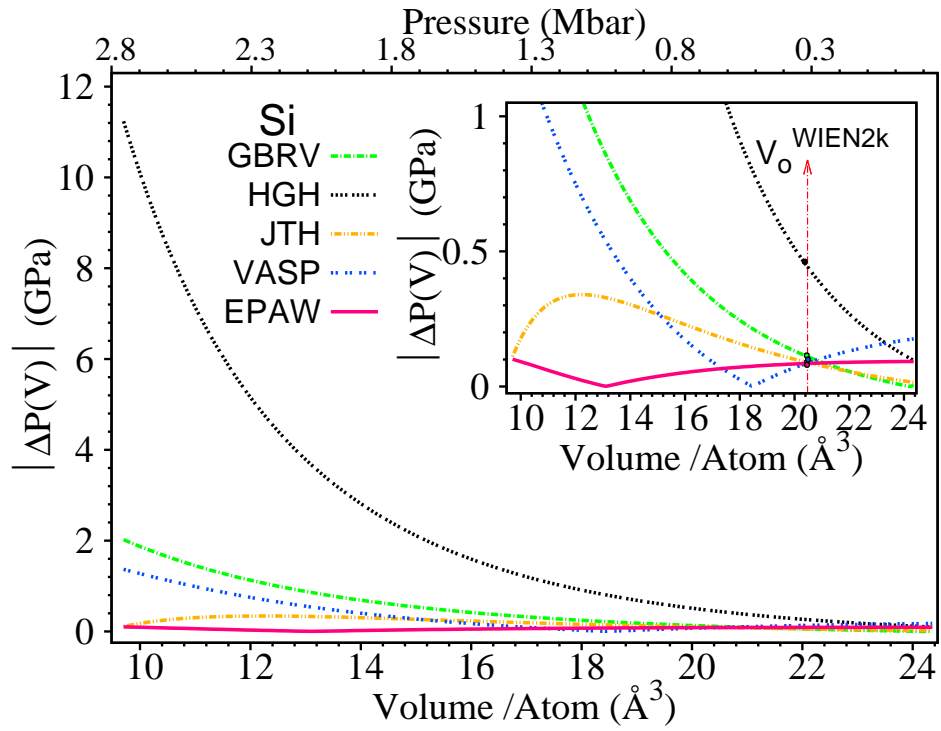}
\caption{Comparison of different atomic data-sets for elemental silicon crystal in the diamond structure. The vertical dotted line represents the equilibrium volume at zero pressure ($V_0$) predicted by WIEN2k code.}\label{SiEOS}
\end{figure}

\begin{figure}[bth!]
\includegraphics[width=0.75\linewidth]{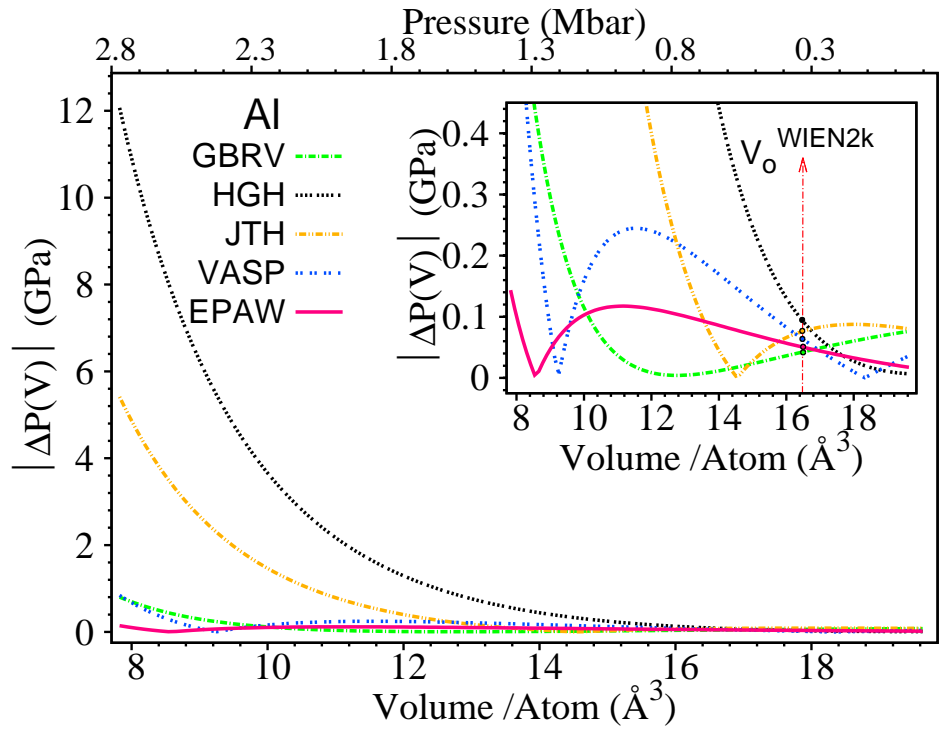}
\caption{Comparison between $\vert \Delta P(V) \vert$ produced by some atomic data-sets for elemental aluminum (fcc) crystal. The vertical dotted line represents the equilibrium volume predicted by WIEN2k code.}\label{AlEOS}
\end{figure}

In the present work, a much larger volume range of $V_{1}=0.475 V_{0}^{AE} \le V \le V_{2}=1.19V_{0}^{AE}$ is chosen. In the example of silicon (Fig. \ref{SiEOS}), GBRV, JTH and VASP data-sets show comparable performances near $V_0$. Increase in pressure causes performance degradation for GBRV and VASP data-sets compared to JTH-PAW performance. However, $\Delta$, $\Delta_1$ and $\Delta_{rel}$ suggest similar performances (Table \ref{tab:table2}) for VASP and JTH data-sets. The case of aluminum (Al) points to an additional issue: the non-uniform data-set performance (see Fig. \ref{AlEOS}). The distance between GBRV, VASP or EPAW from AE-FLAPW results fluctuates throughout the considered pressure range. Since these measures depend only on the area between the AE-FLAPW and the data-set energy-volume curves, they do not reflect the performance fluctuation of these data-sets over the pressure range under consideration. This suggests that goodness measures for an extended pressure range should include also a uniformity criterion.

We define a new class of goodness measures for an atomic data-set with respect to a reference approach as
\begin{equation}
\Delta_{U}(\xi)= A(\Delta \xi) ~ L(\Delta \xi),  \label{obj1}
\end{equation}
where $\xi$ represents either energy ($E$) or pressure ($P$) depending upon the nature of the EoS (energy-volume or pressure-volume) under consideration.  $A(\Delta \xi)$ addresses the closeness of two EoS curves. It is the rescaled area between the AE-FLAPW (AE) and data-set (X) EoS curves between $V_1$ and $V_2$ 
\begin{equation}
A(\Delta \xi) = \frac{1}{V_{2} - V_{1} } \int _{V_{1}}^{V_{2}} \vert \Delta \xi(V) \vert dV, \label{Adef}
\end{equation}
\begin{equation}
\text{where, }\Delta \xi(V) =  \xi^{X}(V) -\xi^{AE}(V), 
\end{equation}
and $\xi^{AE}(V)$ and $\xi^{X}(V)$ are the calculated energy or pressure versus volume curve. $L(\Delta \xi)$ addresses the uniformity of the atomic data-set performance throughout the same compression range. It is the rescaled arc length of the $\Delta \xi(V)$ curve:
\begin{equation}
L(\Delta \xi) =\frac{1}{V_{2} - V_{1} } \int _{V_{1}}^{V_{2}}\sqrt{1+\left(  
\frac{d (\Delta \xi) }{dV} \right)^2 }dV, \label{length}
\end{equation}
with $L(\Delta \xi)\geq 1$.	

\begin{figure}[bth!]
\includegraphics[width=0.75\linewidth]{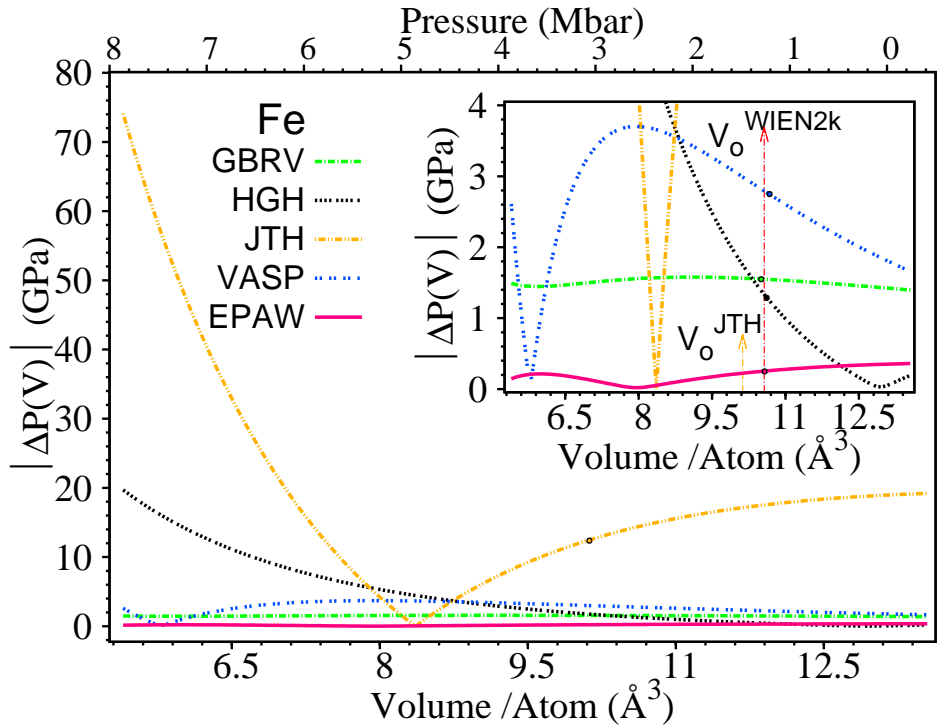}
\caption{ $\vert \Delta P(V) \vert$ curves of different bench-marked atomic data-sets for bcc elemental iron (nonmagnetic) crystal. The vertical dotted lines in the inset represents the equilibrium volume predicted by WIEN2k (green) and HGH (red), while the circles in each curve represents the equilibrium volume corresponding to that scheme.}\label{FeEOS}
\end{figure}

\begin{figure}[bth!]
\begin{tabular}{ll}
\includegraphics[width=0.75\linewidth]{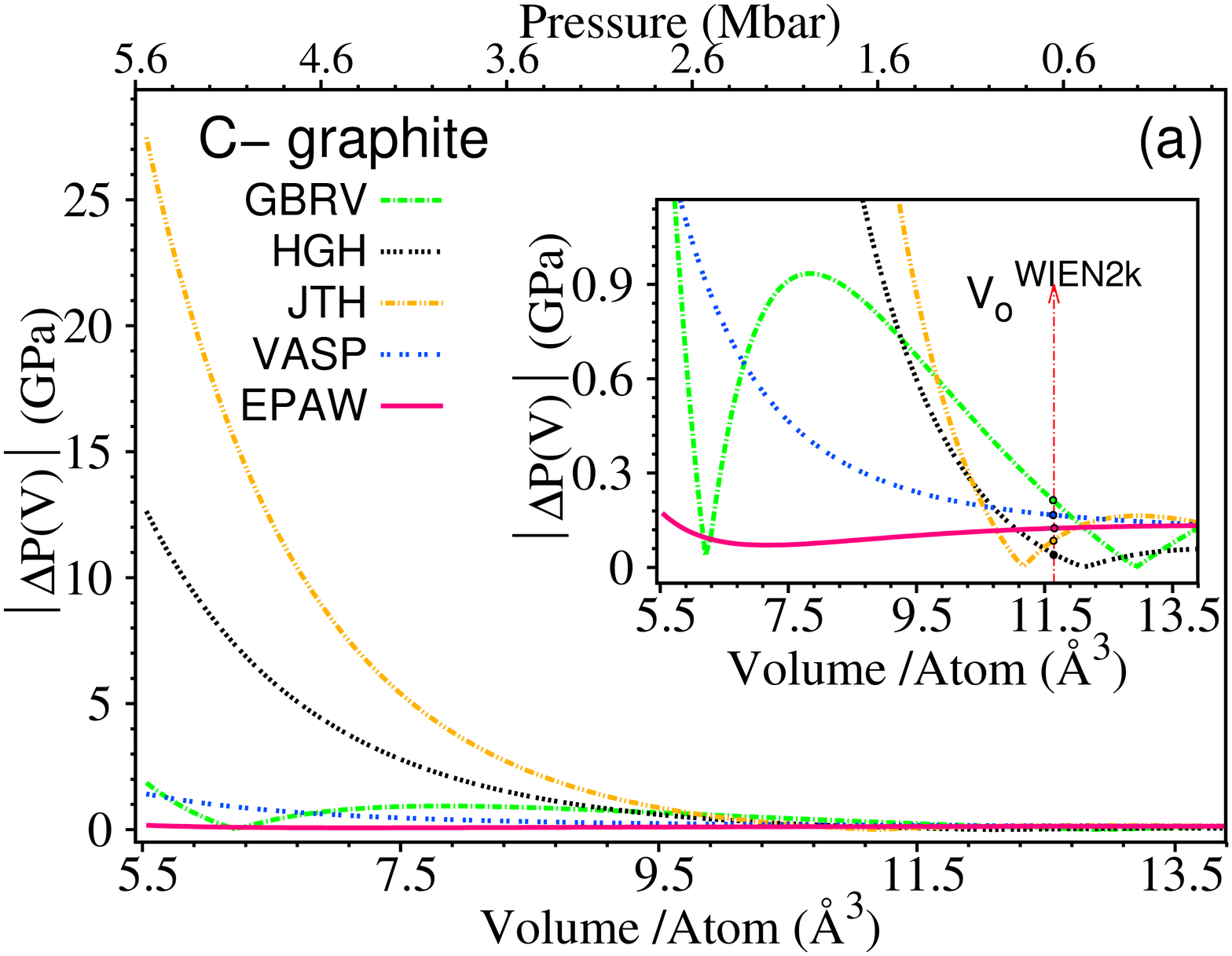}\\
\includegraphics[width=0.75\linewidth]{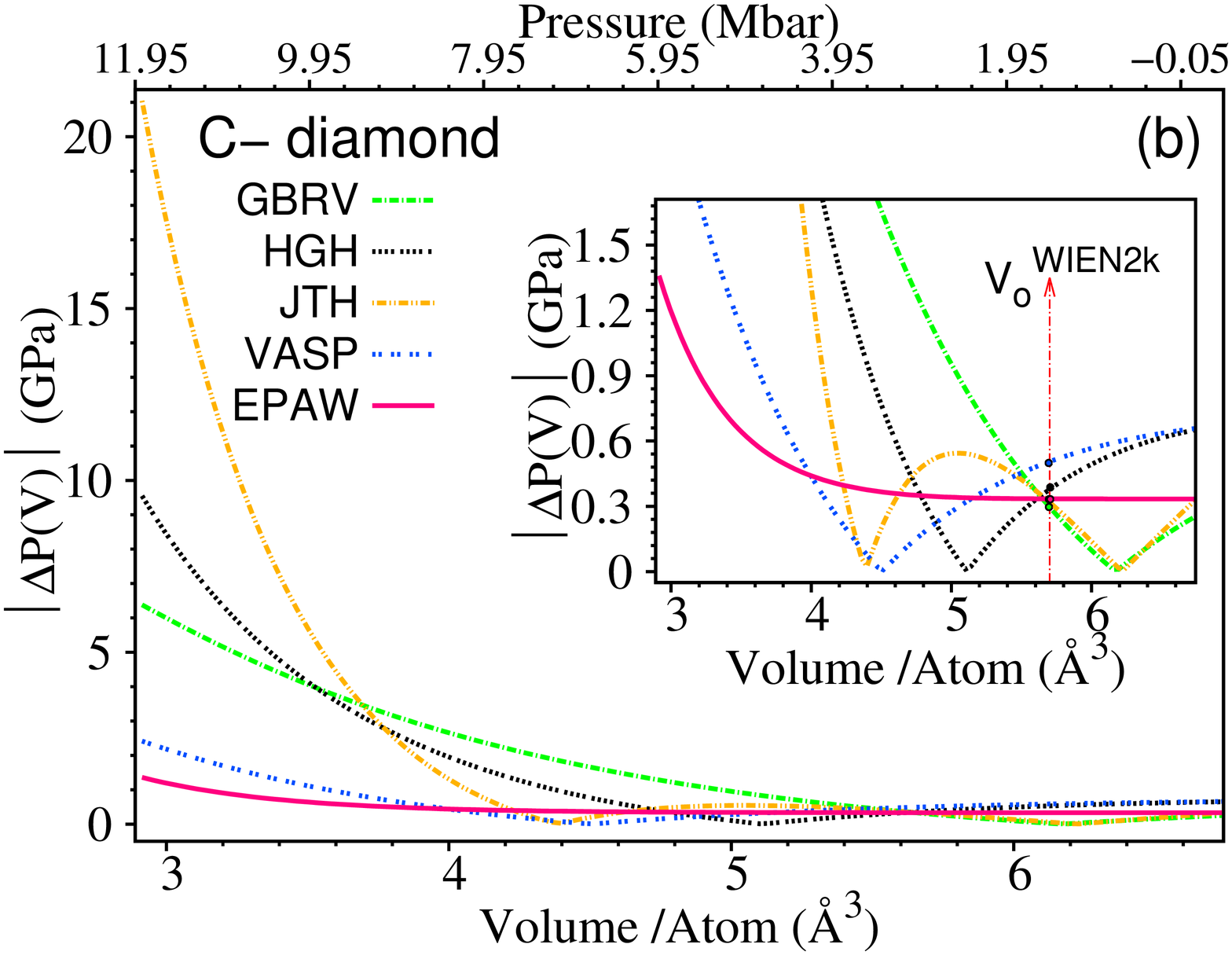}\\
\end{tabular}
\caption{Comparison between $\vert \Delta P(V) \vert$ curves generated by some atomic data-sets for carbon in the (a) graphite and (b) diamond structures. The EPAW data-set generated using the graphite structure also exhibits more uniform performance and outperforms the others for the diamond structure. The vertical dotted line represents the equilibrium volume predicted by the WIEN2k code, while the circles in each curve represents equilibrium volumes for each scheme.  \label{CEOS}}
\end{figure}

\begin{figure}[bth!]
\includegraphics[width=0.75\linewidth]{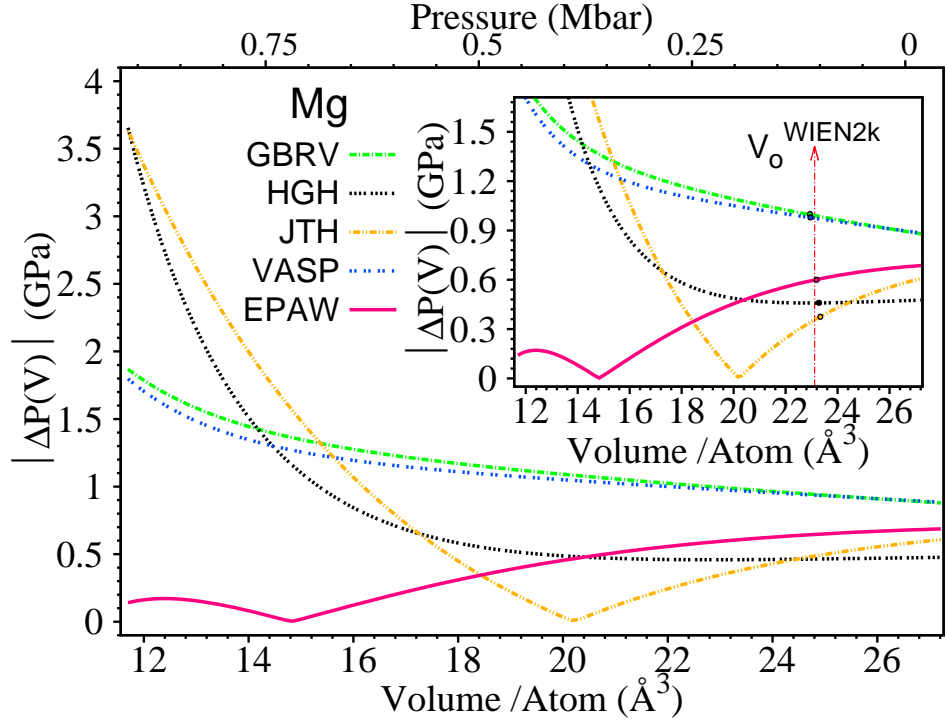}
\caption{Comparison between the performance of several atomic data-sets for an elemental magnesium crystal. The high pressure part has been given a special preference in the PAW data-set (EPAW) optimization. The performance of the optimized EPAW data-sets is better in the high-pressure region. The vertical dotted line represents the equilibrium volume at zero pressure ($V_0$) predicted by WIEN2k code.}\label{MgEOS}
\end{figure}

To calculate $A(\Delta \xi)$ and $L(\Delta \xi)$ in Eqs. \ref{Adef} and \ref{length} we first fit $\xi^{AE} (V)$ and $\xi^{X} (V)$ to a 3$^{rd}$ order finite strain EoS (Eqs. \ref{BM} and/or \ref{PBM}). Goodness measure $\Delta$, $\Delta_{1}$, $\Delta_{rel}$, $\Delta_{U}(E)$ and $\Delta_{U}(P)$ for Fe (Fig. \ref{FeEOS}), C (Fig. \ref{CEOS}), Mg (Fig. \ref{MgEOS}), Si (Fig. \ref{SiEOS}), and Al (Fig. \ref{AlEOS}) are shown in Table \ref{tab:table2}. Except for Mg and Al, all goodness measures indicate that EPAW data-sets perform better than the others. According to $\Delta_{U}(E)$, JTH data-set performs better than EPAW data-set in case of Mg. While for Al, $\Delta$, $\Delta_{1}$, $\Delta_{rel}$ suggest better performance of GBRV than EPAW data-set. However, a closer look at $\vert \Delta P(V) \vert$ curves (Figs. \ref{FeEOS}, \ref{CEOS}, \ref{MgEOS}, \ref{SiEOS}, \ref{AlEOS}) and comparison among the numerical value of these measures indicate that $\Delta_{U}(P)$ is a more sensitive gauge of goodness. In the specific case of silicon, $\Delta$, $\Delta_{1}$ and $\Delta_{rel}$ fail to capture the better  performance of the JTH compared to the VASP data-set (see Fig. \ref{SiEOS}). This is because $A(\Delta E^{\text{VASP}}) \sim A(\Delta E^{\text{JTH}})$ (see Table \ref{tab:table2}). In contrast, $A(\Delta P^{\text{VASP}}) > A(\Delta P^{\text{JTH}})$, correctly pointing to the better performance of the JTH data-set. The use of $A(\Delta P)$ in this case is critical and is also reflected in the new measures $\Delta_{U}(P)$. The non-uniform performance of a particular data-set among a set of data-sets having similar $A(\Delta P)$ is captured by the scaled arc-length $L(\Delta P)$.

 \begin{table*}
\caption{\label{tab:table2}  Relative $\Delta$ values and newly defined goodness measures, $\Delta_{U}(E)$ and $\Delta_{U}(P)$ (Eq. 12), of several atomic data-sets for the bench-marked crystal structures used by Lejaeghere \emph{et al}. \citep{Lejaeghere2014} All the measures are calculated between $V_{1}=0.475 V_{0}^{AE}$ and $V_{2}=1.19V_{0}^{AE}$.}
\begin{ruledtabular}
\begin{scriptsize}
\begin{tabular}{|c|cccccccccc|}
        & Code   & $\Delta  $  & $\Delta_1$  & $\Delta_{rel}$ & $A(\Delta E)$ & $L(\Delta E)$ & $\Delta_U(E)$ & $A(\Delta P)$ & $L(\Delta P)$ & $\Delta_U(P)$ \\ \hline
C       & GBRV      & 10.747   & 13.402   & 0.665         & 0.03207 & 1.00001 & 0.03207 &  0.49686   &  1.02152 &     0.50755        \\        
        & HGH       & 23.031   & 28.687   & 1.433         & 0.05948 & 1.00024 & 0.05949 &  1.94001   &  1.16150 &     2.25332        \\        
        & JTH       & 41.285   & 51.056   & 2.518         & 0.10759 & 1.00105 & 0.10770 &  3.90882   &  1.51659 &     5.92807        \\        
        & VASP      & 6.052    & 7.514    & 0.374         & 0.01694 & 1.00000 & 0.01694 &  0.36263   &  1.00217 &     0.36342        \\        
        & EPAW      & 1.630    & 2.026    & 0.101         & 0.00672 & 1.00000 & 0.00672 &  0.10684   &  1.00004 &     0.10684        \\ \hline
Mg      & GBRV      & 41.856   & 155.055  & 7.150         & 0.06702 & 1.00003 & 0.06703 &  1.17319   &  1.00091 &     1.17426        \\              
        & HGH       & 25.103   & 94.112   & 4.372         & 0.03772 & 1.00002 & 0.03772 &  0.87380   &  1.01870 &     0.89014        \\              
        & JTH       & 18.716   & 68.416   & 3.160         & 0.02289 & 1.00003 & 0.02289 &  0.89576   &  1.01550 &     0.90965        \\             
        & VASP      & 40.129   & 148.438  & 6.851         & 0.06445 & 1.00003 & 0.06445 &  1.12544   &  1.00095 &     1.12650        \\              
        & EPAW      & 14.083   & 51.934   & 2.403         & 0.02497 & 1.00000 & 0.02497 &  0.38523   &  1.00053 &     0.38543        \\ \hline
Al      & GBRV      & 1.187    & 2.631    & 0.117         & 0.00090 & 1.00000 & 0.00090 &  0.09762   &  1.01113 &     0.09871         \\       
        & HGH       & 30.915   & 68.725   & 3.089         & 0.02074 & 1.00022 & 0.02075 &  1.86340   &  1.66816 &     3.10844         \\       
        & JTH       & 10.340   & 22.800   & 1.013         & 0.00666 & 1.00004 & 0.00666 &  0.76005   &  1.22971 &     0.93464         \\       
        & VASP      & 4.230    & 9.382    & 0.417         & 0.00303 & 1.00000 & 0.00303 &  0.15075   &  1.02339 &     0.15427         \\       
        & EPAW      & 2.086    & 4.612    & 0.206         & 0.00175 & 1.00000 & 0.00175 &  0.07103   &  1.00162 &     0.07115         \\ \hline
Si      & GBRV      & 14.237   & 22.425   & 1.059         & 0.02061 & 1.00001 & 0.02061 &  0.52095   &  1.00440 &     0.52325         \\             
        & HGH       & 58.250   & 92.261   & 4.370         & 0.08484 & 1.00025 & 0.08486 &  2.37262   &  1.13720 &     2.69814         \\             
        & JTH       & 6.678    & 10.502   & 0.496         & 0.01010 & 1.00000 & 0.01010 &  0.18340   &  1.00037 &     0.18347         \\             
        & VASP      & 6.564    & 10.361   & 0.489         & 0.00916 & 1.00000 & 0.00916 &  0.33369   &  1.00234 &     0.33447         \\             
        & EPAW      & 1.866    & 2.925    & 0.138         & 0.00316 & 1.00000 & 0.00316 &  0.06256   &  1.00004 &     0.06256         \\ \hline
Fe      & GBRV      & 26.480   & 27.917   & 1.164         & 0.02093 & 1.00004 & 0.02093 &  1.51002   &  1.00127 &     1.51193         \\        
        & HGH       & 80.240   & 84.609   & 3.529         & 0.04538 & 1.00090 & 0.04542 &  4.56263   &  2.80321 &    12.78999         \\        
        & JTH       & 149.258  & 137.160  & 5.898         & 0.12219 & 1.01101 & 0.12354 & 18.78689   & 11.60427 &   218.00814         \\        
        & VASP      & 52.748   & 56.569   & 2.366         & 0.04133 & 1.00015 & 0.04133 &  2.65601   &  1.62376 &     4.31271         \\        
        & EPAW      & 1.757    & 1.874    & 0.078         & 0.00198 & 1.00000 & 0.00198 &  0.20593   &  1.00421 &     0.20679         \\ \hline          
\end{tabular}
\end{scriptsize}
\end{ruledtabular}
\end{table*}

\begin{figure*}
    \centering
    \includegraphics[width=0.49\linewidth]{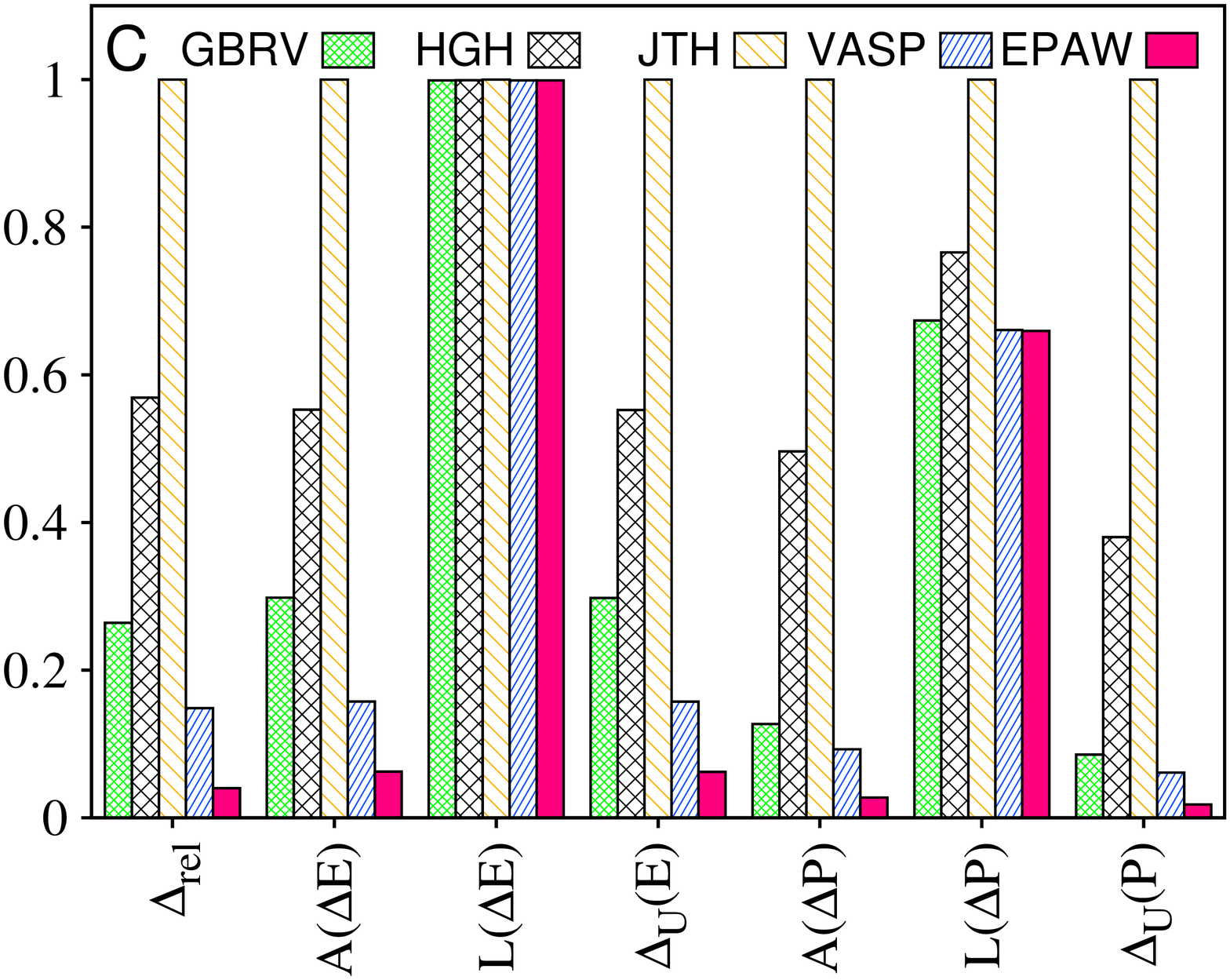}
    \includegraphics[width=0.49\linewidth]{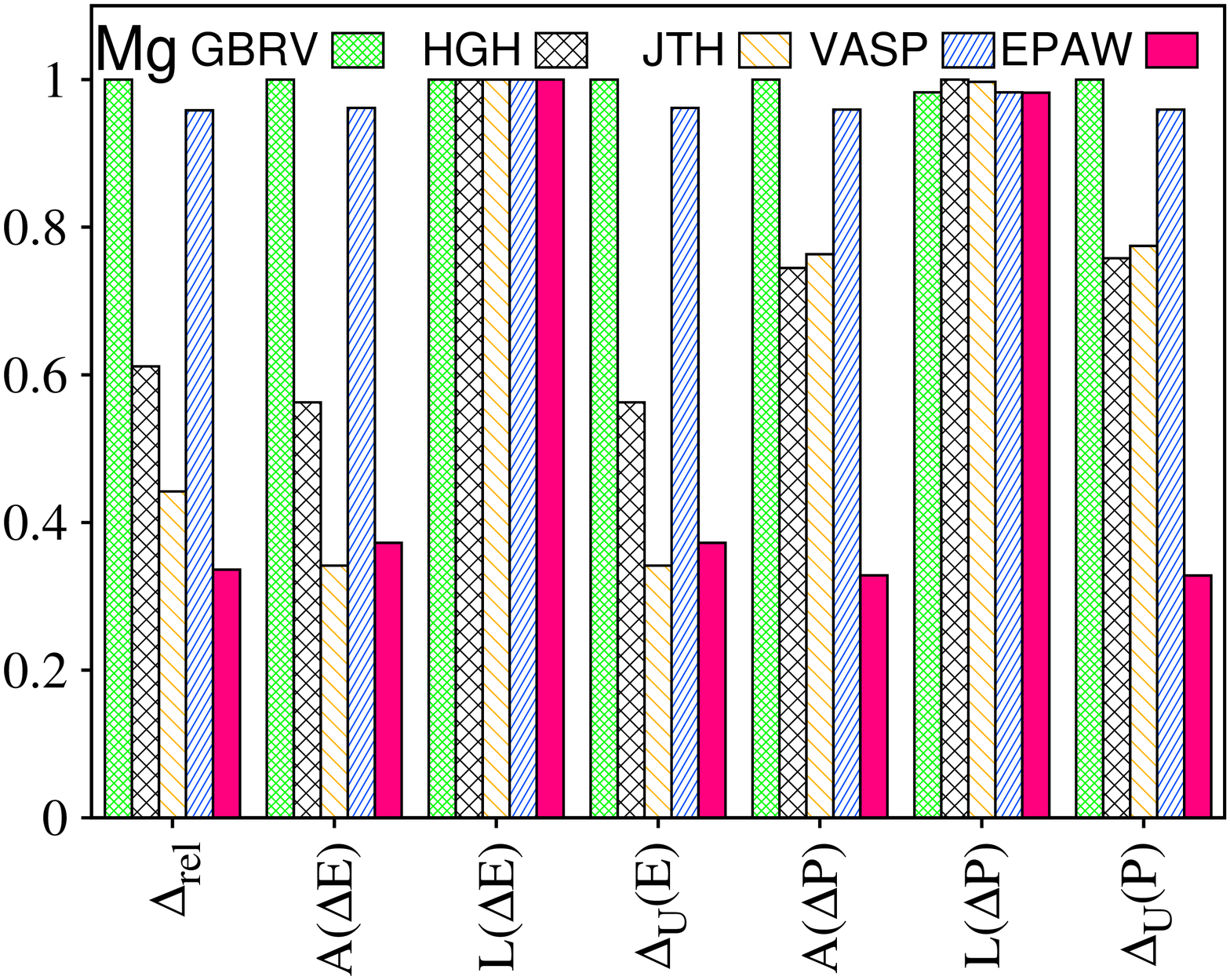}\\
    \includegraphics[width=0.49\linewidth]{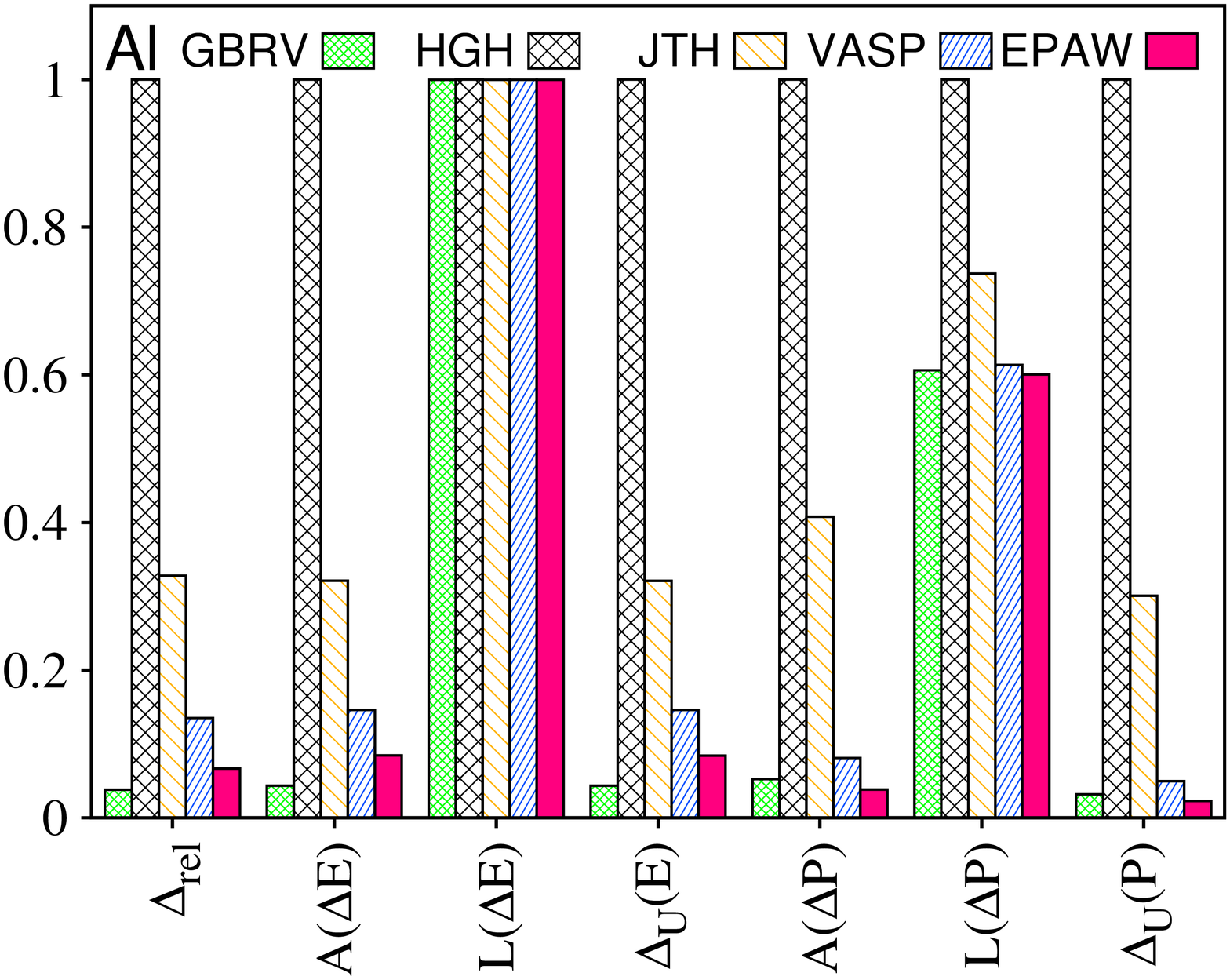}
    \includegraphics[width=0.49\linewidth]{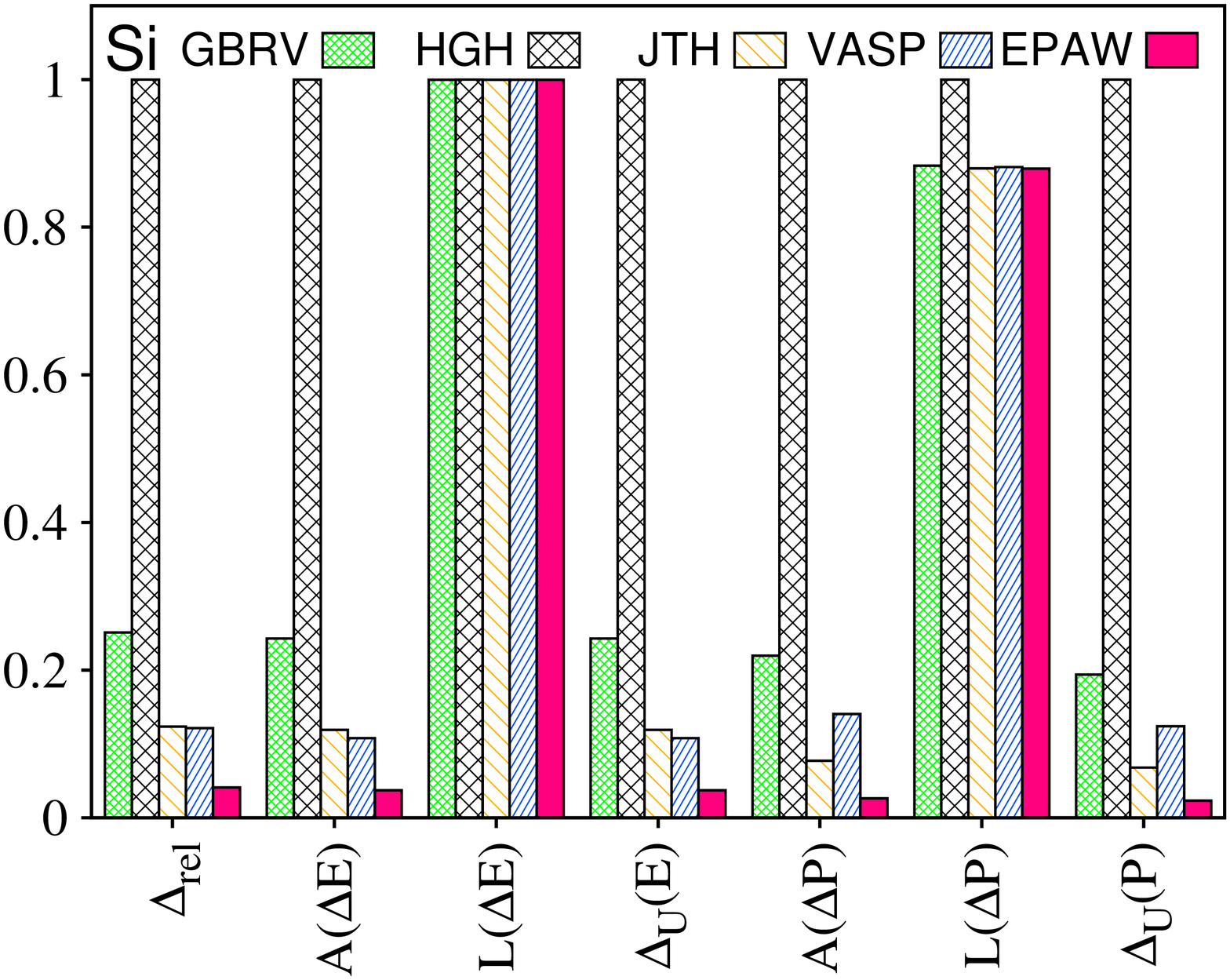}\\
    \includegraphics[width=0.49\linewidth]{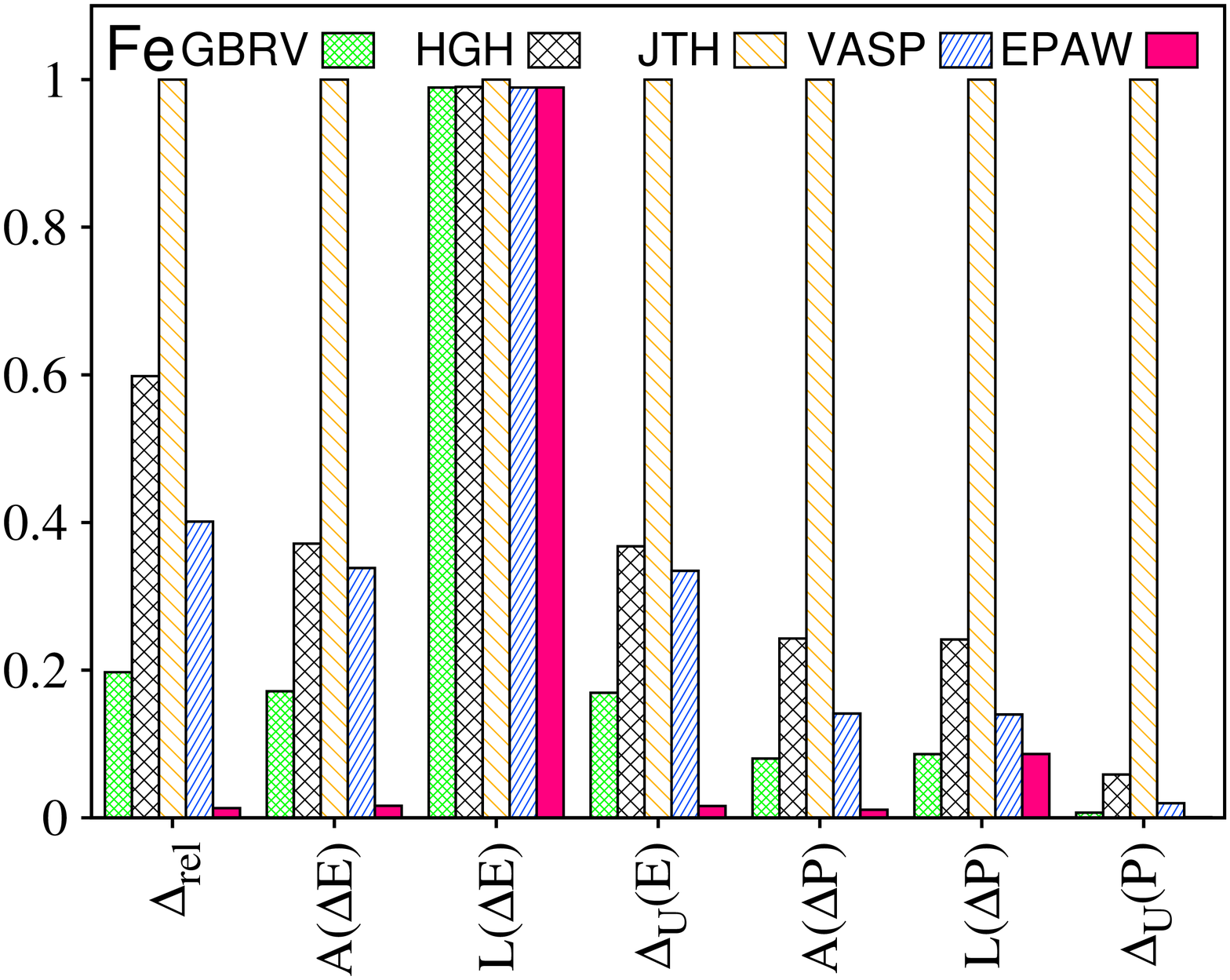}
    \caption{Normalized performance of various data-sets in a wide pressure range according to different evaluation schemes. It visually displays the content of Table \ref{tab:table2} (see text).
}\label{fig:hists}
\end{figure*}

$\Delta_{U}(E)$, and $\Delta_{U}(P)$ along with $A(\Delta E)$, $L(\Delta E)$, $A(\Delta P)$, $L(\Delta P)$ and $\Delta_{rel}$ for C, Mg, Al, Si, and Fe are graphically displayed in Fig. \ref{fig:hists}. This representation captures the contribution by $A(\Delta \xi)$ and $ L(\Delta \xi )$ to the new goodness measures $\Delta_{U}(\xi)$. The heights of different bars (GBRV, HGH, JTH, VASP and EPAW) for a specific measure have been normalized to [0,1]. For carbon and iron all goodness measures ($\Delta_{U}(E)$, $\Delta_{U}(P)$ and $\Delta_{rel}$) rate the data-sets performance in the same order. In contrast, for Mg, Si and Al, $\Delta_{U}(P)$ mirrors more effectively the data-sets performance displayed in Figs. \ref{MgEOS}, \ref{SiEOS} and \ref{AlEOS} respectively. For these elemental solids $\Delta_{U}(E)$ and $\Delta_{rel}$ lead to somewhat different conclusions. This is because both $\Delta_{U}(E)$ and $\Delta_{rel}$ measures are based on energy-volume relation and $L(\Delta E)\sim 1.0$, the latter is not contributing to the $\Delta_{U}(E)$. The difference between the EoS curves is accentuated when pressure is considered instead of energy.

\section{Conclusions} \label{concl}
This communication documents research in the areas of evolutionary computing, electronic structure theory, and their integration to develop an automated recipe for generating uniform, high-quality PAW atomic data-sets throughout an extended pressure range. In addition to minimizing the differences between all electron and PAW atomic logarithmic derivatives, the generation procedure involves the replication of a target all-electron equation of state for elemental solids up to high pressures. These data-sets are generated using similar augmentation radii ($r_c$) and same core-valence states used to generate the standard PAW data-set libraries. Thus, our data-sets are as efficient as those in the standard libraries and yet they reproduce better the all-electron equations of state at high pressures.  

Assessment of the merit of these data-sets for high pressure calculations requires a new goodness evaluation criterion since $\Delta$-gauges frequently used in the extant literature are designed to verify data-set performance around zero pressure. The newly proposed goodness measure argues for using the volume-pressure relation instead of the volume-energy relation over an extended pressure range in assessing the atomic data-set quality. This choice accentuates differences with respect to the target AE-FLAPW equation of state.

The proposed method requires no human supervision once a reliable AE-FLAPW target equation of state for a specific system is defined. We are actively refining as well as extending this work to generate high-quality PAW data-sets for other elements, particularly those abundant in planetary interiors. The present work focuses on improving the accuracy of the results while retaining the computational efficiency of the starting guess. However, this method can also be extended to increase the computational efficiency of the PAW data-sets while retaining their accuracy.

\section{Acknowledgments}
This work is supported primarily by grants NSF/EAR 1348066 and 1503084. Computations are performed at the Minnesota Supercomputing Institute (MSI). NAWH is supported by NSF grant DMR-1507942. Contributions to the ATOMPAW code by Marc Torrent and Fran\c{c}ois Jollet are gratefully acknowledged.

\appendix
\section{PAW variables in the ATOMPAW code}\label{aapndx}

The first step of creating a PAW dataset is to solve the radial Schr\"{o}dinger equation for the self-consistent electronic structure of the atom. Usually one chooses the ground state configuration, although metastable excited state configurations can also be used, as appropriate for the particular electronic structure calculation to be performed. Optionally, a scalar relativistic calculation can be performed in this step, producing the radial function corresponding to the upper component of the Dirac equation averaged over the spin orbit terms. This calculation is based on the formulation by Koelling and Harmon\cite{koelling:1977}. The ATOMPAW code uses subroutines based on the Ultrasoft Pseudopotential (USPP) code \cite{Vand90} modified by Marc Torrent, Fran\c{c}ois Jollet, and N. A. W. Holzwarth. Note that in this formulation, the spin-orbit averaged upper component of the wavefunction is normalized so that the integral of its squared modulus is equal to unity. The lower component wavefunction is not included in the analysis. 

Several pseudo-potential schemes from the literature have been implemented into the ATOMPAW code. These typically depend on the following radial parameters:  
\begin{itemize}
  \item $r_c$ is the radius beyond which all components of the pseudo functions match the all-electron functions. If there are multiple radii specified, this should be the largest radius. If this augmentation sphere radius is the only radius specified, all the other radii listed below are assumed to have the same value.
  \item $r_{shape}$ is the radius which defines the extent of the compensation charge shape function. When using the Kresse form of the PAW formalism,\cite{Kresse1999} such as implemented in the Quantum ESPRESSO code,\cite{PWscf:2009} as opposed to the original Bl\"{o}chl form \cite{Blochl1994}, such as implemented in the Abinit code,\cite{abinit:2009} it is necessary to choose $r_{shape}<r_c$ to properly represent the gradient terms in the exchange-correlation functional.\cite{Torrent20101862}
  \item $r_{vloc}$ is the radius at which the unscreened local pseudopotential vanishes.
  \item $r_{\rm core}$ defines the shape of the smoothed core density $\tilde{n}_{\rm core}(r)$ for $r < r_{\rm core}$ in terms of the all-electron core density function $n_{\rm core}(r)$ and its first few derivatives evaluated at $r_{\rm core}$. 
\end{itemize}

Additionally, it is possible to specify a matching radius for the construction of each the pseudo-partial wave $\tilde{\phi}_{i}(r)$ such that
\begin{equation}
\tilde{\phi}_{i}(r) \equiv \phi_i(r) \;\;\;\;\;{\rm{for ~~~~}} r \ge r_{ci},
\end{equation}
where $r_{ci} \le r_c$ and $\phi_i(r)$ denotes the all-electron partial wave.

The other most common adjustable parameters in the atomic data-sets are the basis set energies $E_{ref1}, E_{ref2}, ...$.  These are chosen separately for each angular momentum channel $l$ in an attempt to make the all-electron and pseudo partial waves as ``complete'' as possible within the augmentation sphere.

\begin{figure*}[bth!]%
\scriptsize
\centering
\begin{tabular}{|lcl|}
\toprule & &\\
C 6                                   & :&  Atomic symbol and number \\
GGA-PBE  loggrid 2001 & :&  XC functional and radial grid specification    \\
\textcolor{black}{2 2 0 0 0 0}        & :&  max. n for each angular momentum: $2s 2p$   \\
\textcolor{black}{2 0 2.0}            & :&  occupation numbers ($2s^2$)    \\
\textcolor{black}{2 1 2.0}            & :&  occupation numbers ($2p^2$)    \\
0 0 0                                 & :&  ends occupation section        \\
\textcolor{black}{c}                  & :&  $1s$ treated as frozen core    \\
\textcolor{black}{v}                  & :&  $2s$ treated as valence        \\
\textcolor{black}{v}                  & :&  $2p$ treated as valence        \\
\textcolor{black}{1}                  & :&  max. $l$ for basis  and projector functions  \\
\textcolor{red}{\textbf{1.3 1.1 1.3 1.3}}  & :&  radial values (Bohr) ($r_c$, $r_{\rm shape}$, $r_{\rm vloc}$, $r_{\rm core}$)  \\
\textcolor{black}{y}                  & :&  additional basis function for $l=0$     \\
\textcolor{red}{\textbf{16.0}}        & :&  $E_{ref1}$ energy (in Ry) for additional $l=0$ basis function    \\
\textcolor{black}{n}                  & :&  no further $l=0$ basis functions \\
\textcolor{black}{y}                  & :&  additional basis function for $l=1$     \\
\textcolor{red}{\textbf{12.0}}        & :&  $E_{ref2}$ energy (in Ry) for additional $l=1$ basis function  \\  
\textcolor{black}{n}                  & :&  no further $l=1$ basis functions \\
\textcolor{blue}{MODRRKJ   VANDERBILTORTHO     BESSELSHAPE}  & :&  generation scheme for PS partial waves and projectors \\
\textcolor{blue}{2 0.0 MTROULLIER }   & :&  local pseudopotential parameters ($l_{loc}=2$, $E_{loc}=0$ Ry)  \\
\textcolor{red}{\textbf{1.3}}         & :&  $r_{c1}$ matching radius for first  s partial wave  \\
\textcolor{red}{\textbf{1.3}}         & :&  $r_{c2}$ matching radius for second s partial wave  \\
\textcolor{red}{\textbf{1.3}}         & :&  $r_{c3}$ matching radius for first  p partial wave  \\
\textcolor{red}{\textbf{1.3}}         & :&  $r_{c4}$ matching radius for second p partial wave  \\
XMLOUT                                & :& create data-set for ABINIT in xml format             \\
default                               &  &                                                      \\
PWSCFOUT                              & :& create data-set for Quantum ESPRESSO \\
UPFDX 0.0125d0 UPFXMIN -7.d0 UPFZMESH 6.d0 & :& UPF grid parameters \\
END        & &                                         \\& & \\
\toprule 
\end{tabular}
\caption{An example input file for C with descriptive commentary added.
  \label{fig:table}}
\end{figure*}

\end{document}